\documentclass[12pt,a4paper]{article}
\usepackage{graphicx}
\usepackage{dcolumn}
\usepackage{bm}
\usepackage{amsmath,amsfonts,amssymb,mathrsfs}
\usepackage[dvipsnames]{xcolor}
\usepackage{bbm}
\usepackage{braket,xcolor}
\usepackage{verbatim}
\usepackage{subcaption}
\usepackage[normalem]{ulem}
\usepackage{multirow}
\usepackage{amsfonts,mathtools,resizegather}
\usepackage[utf8]{inputenc}
\usepackage{scalerel}
\usepackage{makecell}
\usepackage{amsfonts, mathtools,resizegather}
\usepackage{multicol}
\usepackage[dvipsnames]{xcolor}
\usepackage[bb=boondox]{mathalfa}
\usepackage[export]{adjustbox}
\usepackage{caption,jheppub}

\usepackage{xcolor}
\usepackage{graphicx, physics}
\usepackage{amsmath}
\usepackage{amssymb}
\usepackage{amsbsy}
\usepackage{amsthm}
\usepackage{amsfonts}
\usepackage{upgreek}
\usepackage{bbm}
\usepackage{amsfonts}
\usepackage{slashed}
\usepackage{algorithm}
\usepackage{algpseudocode}
\usepackage{orcidlink}

\title{\Large Analytic spread complexity in three-level systems and the spacing-ratio distribution}

\author[a,b]{Pratik Nandy\,\orcidlink{0000-0001-5383-2458}}

\emailAdd{pratik.nandy@vub.be}

\affiliation[a]{Theoretische Natuurkunde, Vrije Universiteit Brussel \& International Solvay Institutes,\\ Pleinlaan 2, B-1050 Brussels, Belgium}
\affiliation[b]{Division of Fundamental Mathematical Science,\\ RIKEN Center for Interdisciplinary Theoretical and Mathematical Sciences (iTHEMS),\\ Wako, Saitama 351-0198, Japan}

\abstract{The spread complexity characterizes the spreading of a quantum state over the Krylov basis and has been proposed to capture the correlations between energy levels. How such correlations are reflected in the dynamics has been studied mainly numerically for large matrices. Here we consider three-level systems, the smallest setting with spacing correlations, and derive exact results for the spread complexity, the Krylov variance, and the Krylov entropy. At infinite temperature, the Krylov basis depends on the spectrum only through the ratio of consecutive level spacings, and the spectral variance enters only as an overall time scale. For Gaussian ensembles, these variables are statistically independent, and the averaged spread complexity and variance reduce to one-dimensional integrals over the spacing-ratio distribution encoding the level statistics. We establish universal late-time values of the complexity and variance, and an algebraic relaxation with exponents set by the Dyson index. We also outline the extension to $d$ levels, with the ensemble-averaged spread complexity expressed as a $(d-2)$-dimensional integral over the joint distribution of consecutive spacing ratios.}

\begin{document}

\maketitle
\flushbottom


\section{Introduction}

Since the work of Wigner, Dyson and Mehta \cite{Wigner1, dyson1962statisticalI, MEHTA1960} and the Bohigas–Giannoni–Schmit conjecture \cite{Bohigas:1983er}, energy-level statistics have become a standard tool for characterizing quantum chaos, both in systems with a chaotic classical limit and in many-body systems without a classical counterpart \cite{haake2018}. Spectra of chaotic systems exhibit level repulsion and follow the local statistics of Gaussian random-matrix ensembles \cite{mehta1991random}, and integrable systems typically have uncorrelated energy levels \cite{Berry:1977tabor}. Complementary to level statistics, dynamical probes such as the spectral form factor (SFF) \cite{Guhr:1997ve, Cotler:2016fpe}, operator growth in the Heisenberg picture \cite{Roberts:2018operator, Parker:2018yvk}, and the spreading of a state over the Hilbert space \cite{Balasubramanian:2022tpr} characterize the evolution of states and operators. The SFF is determined by the spectrum alone, but operator growth and the spreading of states also depend on the initial operator or state. In general, it is not straightforward to relate such dynamical quantities quantitatively to the statistics of energy levels.
 
In this paper, we relate the correlations between energy levels quantitatively to a class of dynamical quantities in the Krylov basis \cite{Nandy:2024evd, Baiguera:2025dkc, Rabinovici:2025otw}. The Lanczos algorithm \cite{lanczos1950iteration,viswanath1994recursion} maps Hamiltonian evolution of quantum states onto a tight-binding \emph{Krylov chain}, whose hopping amplitudes (Lanczos coefficients) are determined by the spectral decomposition of the initial state. Their response to certain deformations of the Hamiltonian or the initial state is governed by Toda flows \cite{Dymarsky:2019elm, Angelinos:2025drf, Takahashi:2025iol}.
The time-evolved state defines a probability distribution on the Krylov chain. Its mean defines the Krylov complexity for operators \cite{Parker:2018yvk} and the spread complexity for states \cite{Balasubramanian:2022tpr}, and its Shannon entropy provides the Krylov entropy \cite{Barbon:2019wsy}. 
The variance quantifies the width of this distribution, while its higher cumulants \cite{Bhattacharjee:2022ave} characterize fluctuations beyond the mean and have been studied for systems driven across a quantum phase transition \cite{Grabarits:2025xys}. In finite-dimensional Hilbert spaces, the chain has finite length, and these quantities saturate at late times \cite{Barbon:2019wsy}. The spread complexity has a characteristic temporal profile, which has been studied for the thermofield-double (TFD) state and for generic initial states \cite{Balasubramanian:2022tpr, Erdmenger:2023wjg, Bhattacharjee:2022qjw}. For chaotic systems, it grows and exhibits a peak, which has been proposed as a signature of chaotic spectra \cite{Balasubramanian:2022tpr}. The peak is usually absent for integrable systems \cite{Balasubramanian:2022tpr}, and its height has been used to probe integrable-to-chaotic transitions \cite{Bhattacharjee:2024yxj}. Finally, it saturates at a plateau that depends on the initial state and on the dimension of the Hilbert space. So far, the connection between these features and correlations between energy levels has been established mainly numerically, using the Lanczos algorithm or the Hessenberg decomposition. Krylov methods have also been extended to time-dependent generators \cite{Takahashi:2024hex}.
 
Exact Krylov dynamics is known for Hamiltonians built from the generators of a Lie algebra, such as $SL(2,\mathbb{R})$, $SU(2)$, or the Heisenberg--Weyl algebra. In these cases, the Lanczos coefficients are fixed by the algebra, and the spread complexity of coherent states is obtained in closed form or approximate closed form for any dimension \cite{Caputa:2021sib, Balasubramanian:2022tpr, Bhattacharjee:2022qjw,  Balasubramanian:2025xkj}. The corresponding spectra are equally spaced or otherwise constrained by the symmetry, and carry no information about level statistics. For generic spectra, the Lanczos coefficients are functionals of the energy levels. In random-matrix ensembles, the Krylov quantities fluctuate across realizations, and their averages depend on the joint distribution of the energy levels. Analytic control over these averages has so far been limited to two-level systems \cite{Caputa:2024vrn, Nandy:2024mml}, and the statistics of the Lanczos coefficients are known only for large-dimensional spectra in a few specialized cases \cite{Balasubramanian:2022tpr, Erdmenger:2023wjg, Balasubramanian:2022dnj, Balasubramanian:2023kwd, Nandy:2024zcd, Qu:2025lgo}. For two levels, the spectrum consists of a single gap, and averages reduce to integrals over the distribution of that gap. For the Gaussian $\beta$-ensembles \cite{Dumitriu:2002beta}, this distribution is the Wigner surmise, and the averaged spread complexity has been obtained in closed form for arbitrary Dyson index in terms of a confluent hypergeometric function \cite{Nandy:2024mml}. In this case, however, the Lanczos coefficients cannot encode correlations between different gaps, which are a defining feature of random-matrix spectra. For large spectra, analytic results are asymptotic in the dimension or rely on an approximate form of the Krylov kernel that is only justified numerically \cite{Basu:2026gvl}. An exact analytic treatment of generic spectra that retains nontrivial correlations between energy levels is still lacking. It is also not known to what extent Krylov dynamics is sensitive to level repulsion and to correlations between adjacent spacings.

Here we address this question for the spreading of states in three-level systems, the smallest systems in which correlations between level spacings enter the Krylov dynamics. At infinite temperature, the Krylov dynamics depends on the spectrum only through the ratio of consecutive spacings, so that its ensemble averages are controlled by the spacing-ratio distribution. We show that the Krylov basis can be obtained in closed form for arbitrary energy gaps and spectral weights, without assuming any algebraic structure of the spectrum. We consequently obtain explicit expressions for the spread complexity $K_S (t)$, the variance $V_S (t)$ of the Krylov distribution, and the Krylov entropy $E_S (t)$. In Gaussian random-matrix ensembles, these results yield exact ensemble averages of Krylov-space quantities beyond the two-level case. The full time dependence of the averaged spread complexity, including its early growth, its peak and its approach to the late-time value, is obtained analytically and compared with numerical simulations.

In particular, we obtain the closed-form expression for the ensemble-averaged spread complexity of three-level systems at infinite temperature, for the Gaussian orthogonal (GOE), unitary (GUE), and symplectic (GSE) ensembles and for independent and identically distributed (IID) levels:
\begin{align}
K_S(t)=1-\frac{2}{9}\int_0^1\mathrm{d} r\,\frac{P_{\beta_D}(r)}{1+r+r^2}\sum_{f\in\{1,\,r,\,1+r\}}\alpha_f \;{}_1F_1\Big(1+\frac{3\beta_D}{2};\frac{1}{2};-\frac{f^2t^2}{4\kappa_{\beta_D} (1+r+r^2)}\Big)\,.
\label{eq:main}
\end{align}
Here, $\beta_D$ is the Dyson index ($\beta_D = 0, 1,2,4$ for IID levels, GOE, GUE and GSE respectively), $P_{\beta_D}(r)\propto(r+r^2)^{\beta_D}/(1+r+r^2)^{1+3\beta_D/2}$ is the distribution of the ratio $r$ of consecutive level spacings \cite{Atas2013distribution}, $\alpha_f$ are quadratic polynomials in $r$, and $\kappa_{\beta_D}=\beta_D/2$ for the Gaussian ensembles and $\kappa_0=1/3$ for IID levels, with the Gaussian weights chosen as in Sec.\,\ref{sec:jpdf}. The variance has the same structure, and the Krylov entropy is given by a two-dimensional integral \eqref{eq:SKavg}. Equation \eqref{eq:main}, obtained for three-level systems, qualitatively agrees with
the behavior found numerically for large random matrices. The average grows quadratically at early times, rises above its late-time value for the Gaussian ensembles, with a peak whose height increases with $\beta_D$, and shows no peak for IID levels.

At infinite temperature, the Krylov basis is obtained from a fixed frame by a rotation through an angle $\theta$, which encodes the ratio of consecutive spacings. The overall scale $Q$ of the spectrum, proportional to the variance of the levels, only controls the time scale. In the Gaussian ensembles, $Q$ and $\theta$ are statistically independent, and the distribution of $\theta$ coincides with the distribution of the spacing ratio derived in Ref.\,\cite{Atas2013distribution} from $3\times3$ random matrices, a widely used diagnostic of level correlations (see also Ref.\,\cite{Oganesyan:2007wpd}). The average over the scale produces the confluent hypergeometric functions in Eq.\,\eqref{eq:main}. At infinite temperature, the late-time values $K_S^{\infty}=1$, $\delta K_S^2=\frac{11}{54}$ and $V_S^{\infty}=\frac{25}{54}$ are the same for every three-level spectrum. The late-time value of $E_S$, on the other hand, depends on $\theta$ and is invariant under a duality that exchanges nearly degenerate and nearly equally spaced configurations, $E_S^{\infty}(y)=E_S^{\infty}(\sqrt{1-y^2})$ with $y=|\sin3\theta|$.
 
We determine how the averages approach their late-time values. For the complexity and the variance, the decay is algebraic in every ensemble, with the exponents fixed by the Dyson index $\beta_D$. For IID levels and the GOE, the decay is $t^{-2}$ in both cases. We further analyze the low-temperature behavior, which is dominated by nearly degenerate ground-state doublets and yields closed-form limits for the ratios $V_S^{\infty}/K_S^{\infty}$ and $E_S^{\infty}/K_S^{\infty}$ in terms of Dirichlet eta functions.
 
The paper is organized as follows. In Sec.\,\ref{sec:setup}, we define the quantities under study and work out the two-level case. In Sec.\,\ref{sec:exact}, we express the Krylov wavefunctions and their late-time averages through the spectral weights and construct the Krylov basis for three levels. Section \ref{sec:shape} discusses the infinite-temperature state, and Sec.\,\ref{sec:rmt} the Gaussian ensembles, including finite temperature and the approach to the late-time values. We summarize our conclusions in Sec.\,\ref{sec:disc} and outline how the construction extends to larger spectra.
 
\section{Krylov-space quantities for a finite spectrum}
\label{sec:setup}
 
\subsection{Lanczos algorithm and the Krylov distribution}
 
Let $H=\sum_{i=1}^d E_i\ket{E_i}\bra{E_i}$ be a Hamiltonian on a $d$-dimensional Hilbert space and $\ket{\Psi(0)}=\sum_ic_i\ket{E_i}$ a normalized initial state. The Lanczos algorithm \cite{lanczos1950iteration} constructs an orthonormal basis $\{\ket{K_n}\}$ of the Krylov space spanned by $H^n\ket{\Psi(0)}$. Starting from $\ket{K_0}=\ket{\Psi(0)}$, the recursion \cite{viswanath1994recursion}
\begin{align}
H\ket{K_n}=a_n\ket{K_n}+b_{n+1}\ket{K_{n+1}}+b_n\ket{K_{n-1}}\,,\qquad b_0=0\,,
\label{eq:lanczos}
\end{align}
defines the Lanczos coefficients $a_n$ and $b_n>0$ and terminates when $b_{\mathcal{K}}=0$, where $\mathcal{K}\le d$ is the dimension of the Krylov space. Expanding the time-evolved state in the Krylov basis, we have \cite{Balasubramanian:2022tpr}
\begin{align}
\ket{\Psi(t)}=e^{-iHt}\ket{\Psi(0)}=\sum_{n=0}^{\mathcal{K}-1}\phi_n(t)\ket{K_n}\,,~~~ \phi_n(t)=\langle K_n|\Psi(t)\rangle\,.
\label{eq:expand}
\end{align}
The amplitudes, or the Krylov wavefunctions $\phi_n(t)$, obey the three-term recurrence relation $i\dot\phi_n(t)=a_n\phi_n(t)+b_n\phi_{n-1}(t)+b_{n+1}\phi_{n+1}(t)$ with $\phi_n(0)=\delta_{n0}$. Since the evolution is unitary, the probabilities $p_n(t)=|\phi_n(t)|^2$ define the Krylov distribution and sum to unity. We consider its mean (spread complexity), its variance (Krylov variance), and its Shannon entropy (Krylov entropy):
\begin{align}
K_S(t)=\sum_nn\,p_n(t)\,,~~ V_S(t)=\sum_nn^2p_n(t)-K_S(t)^2\,,~~ E_S(t)=-\sum_np_n(t)\ln p_n(t)\,.
\label{eq:defs}
\end{align}
The late-time averages are denoted by a superscript, $X^\infty=\lim_{T\to\infty}T^{-1}\int_0^T\mathrm{d} t\,X(t)$. We also use the temporal variance of the complexity,
\begin{align}
\delta K_S^2=\lim_{T\to\infty}T^{-1}\int_0^T\mathrm{d} t\,\big[K_S(t)-K_S^\infty\big]^2\,,
\label{eq:dK}
\end{align}
which measures the size of the fluctuations of $K_S(t)$ around its late-time value $K_S^\infty$.
 
\subsection{Return amplitude and spectral weights}
 
The Lanczos coefficients and the amplitudes are determined by the return amplitude (also called the survival amplitude)
\begin{align}
\mathcal{R}(t)=\bra{\Psi(0)}e^{-iHt}\ket{\Psi(0)}=\sum_kw_k\,e^{-iE_kt}\,,\qquad w_k=|c_k|^2=|\langle E_k|\Psi(0)\rangle|^2\,,
\label{eq:measureS}
\end{align}
and hence by the energy levels $E_k$ and the populations $w_k$ in the initial state, which we refer to as the spectral weights. Levels with $w_k=0$ do not contribute to $\mathcal{R}(t)$, and a set of degenerate levels enters as a single level with the sum of their weights. Hence $\mathcal{K}$ equals the number of distinct energies with non-zero weight. We mainly consider the TFD-type state (coherent Gibbs state)
\begin{align}
\ket{\psi_\beta}=\frac{1}{\sqrt Z}\sum_{k=1}^d e^{-\beta E_k/2}\ket{E_k}\,,\qquad Z=\sum_{k=1}^d e^{-\beta E_k}\,,
\label{eq:tfd}
\end{align}
which has weights $w_k=e^{-\beta E_k}/Z$. The TFD state in the doubled Hilbert space, evolved with $H\otimes\mathbb{1}$, has the same return amplitude and the same Krylov distribution.
 
\subsection{The two-level example}
\label{sec:N2}
 
For $d=2$, the Krylov construction can be carried out explicitly, and the result serves as a reference for what follows. Let $\omega=E_2-E_1$ and $\ket{K_0}\equiv\ket{\Psi(0)}=c_1\ket{E_1}+c_2\ket{E_2}$ with $w_1+w_2=1$. The first Lanczos coefficient is given by the mean energy
\begin{align}
a_0=\bra{K_0}H\ket{K_0}=w_1E_1+w_2E_2\equiv\bar E\,,~~~ E_1-\bar E=-w_2\,\omega\,,~~~ E_2-\bar E=w_1\,\omega\,.
\end{align}
The recursion \eqref{eq:lanczos} with $n=0$ gives $b_1\ket{K_1}=(H-a_0)\ket{K_0}=-c_1w_2\,\omega\ket{E_1}+c_2w_1\,\omega\ket{E_2}$. Since $\ket{K_1}$ is normalized, the norm of the right-hand side fixes $b_1$, and its direction gives the second Krylov vector:
\begin{align}
b_1^2=\omega^2\big(w_1w_2^2+w_2w_1^2\big)=w_1w_2\,\omega^2\,,\qquad
\ket{K_1}=\frac{-c_1w_2\ket{E_1}+c_2w_1\ket{E_2}}{\sqrt{w_1w_2}}\,.
\label{eq:N2K1}
\end{align}
The Krylov space is two-dimensional, $a_1=\bra{K_1}H\ket{K_1}=w_2E_1+w_1E_2$, and $b_2=0$. The amplitude on the second site is
\begin{align}
\phi_1(t)=\bra{K_1}e^{-iHt}\ket{\Psi(0)}=\sqrt{w_1w_2}\,\big(e^{-iE_2t}-e^{-iE_1t}\big)\,.
\end{align}
The probability on the second site is simply
\begin{align}
p_1(t)=|\phi_1(t)|^2=4w_1w_2\sin^2(\omega t/2)\equiv P(t)\,, \label{prob}
\end{align}
so that $p_0(t)=|\phi_0(t)|^2=1-P(t)$. The complexity, variance, and entropy are given by
\begin{gather}
K_S(t)=P(t)\,,~~V_S(t)=P(t)(1-P(t))\,,~~ E_S(t)=-P(t)\ln P(t)-(1-P(t))\ln(1-P(t))\,.
\label{eq:N2S}
\end{gather}
For the coherent Gibbs state \eqref{eq:tfd}, $w_1w_2=\frac14\mathrm{sech}^2(\beta\omega/2)$. At $\beta=0$, one has $K_S=\sin^2(\omega t/2)$ and $V_S=\frac14\sin^2\omega t$, with late-time values $K_S^\infty=\frac12$, $V_S^\infty=\frac18$ and $E_S^\infty=2\ln2-1$ (see Eq.\,\eqref{eq:s2level} for general $\beta$). All three quantities depend on the spectrum only through $\omega t$, so that their ensemble averages are integrals over the distribution of the single gap $\omega$ \cite{Nandy:2024mml}.
 
\section{Krylov basis of a finite spectrum}
\label{sec:exact}
 
\subsection{General discrete spectrum}
 
Consider $d$ distinct levels $E_k$ with weights $w_k>0$, $\sum_kw_k=1$, so that $\mathcal{K}=d$. Since only the weights $w_k=|c_k|^2$ enter the return amplitude \eqref{eq:measureS}, the phases of $c_k=\sqrt{w_k}\,e^{i\varphi_k}$ can be absorbed into the energy eigenstates, $\ket{E_k}\to e^{i\varphi_k}\ket{E_k}$, and we take $c_k=\sqrt{w_k}$ without loss of generality. The initial state is then
\begin{align}
\ket{K_0}=\sum_k\sqrt{w_k}\,\ket{E_k}\,.
\label{k0}
\end{align}
The Krylov vectors can be expressed in the energy basis as
\begin{align}
\ket{K_n}=\sum_ku_{n,k}\ket{E_k}\,,~~~ u_{n,k}=\langle E_k|K_n\rangle\,,~~~u_{0,k} = \sqrt{w_k}\,.
\end{align}
Since $H$ and $\ket{K_0}$ are real in this basis, the recursion \eqref{eq:lanczos} produces real components $u_{n,k}$. Since the Krylov vectors are orthonormal, $\langle K_m|K_n\rangle=\delta_{mn}$, we have $\sum_ku_{m,k}u_{n,k}=\delta_{mn}$. Thus, the components $u_{n,k}$ form an orthogonal matrix, and the Krylov wavefunctions are given by
\begin{align}
\phi_n(t)=\langle K_n|\Psi(t)\rangle=\bra{K_n}e^{-iHt}\ket{K_0}=\sum_{k=1}^d \sqrt{w_k}\,u_{n,k}\,e^{-iE_kt}\,,
\label{eq:amp}
\end{align}
where we have used Eq.\,\eqref{k0} with $\ket{K_0}=\ket{\Psi(0)}$.
 
The late-time averages follow from Eq.\,\eqref{eq:amp}. We write
\begin{align}
K_S(t)=\sum_{k,l}\sqrt{w_kw_l}\,X_{kl}\,e^{-i(E_k-E_l)t}\,,\qquad X_{kl}=\sum_nn\,u_{n,k}u_{n,l}\,,
\label{eq:Xdef}
\end{align}
where $X$ is the position operator on the Krylov chain expressed in the energy basis. Assuming that no two differences $E_k-E_l$ with $k\neq l$ coincide, one finds the identities:
\begin{gather}
\overline{p_n}\equiv p_n^{\infty}=\sum_kw_k\,u_{n,k}^2\,,\qquad K_S^{\infty}=\sum_kw_kX_{kk}\,,\qquad \delta K_S^2=\sum_{k\neq l}w_kw_l\,X_{kl}^2\,,\nonumber\\
V_S^{\infty}=\sum_kw_k\,(X^2)_{kk}-\big(K_S^{\infty}\big)^2-\delta K_S^2\,.
\label{eq:timeavg}
\end{gather}
For equal weights, $w_k=1/d$, the orthogonality of $u_{n,k}$ gives $p_n^{\infty}=1/d$ for every $n$, and hence
\begin{align}
    K_S^{\infty}=\frac{d-1}{2}\,,~~~~~ V_S^{\infty}+\delta K_S^2=\frac{d^2-1}{12}\,, \label{llt}
\end{align}
independently of the levels. The late-time variance is thus bounded by the variance of the uniform distribution on the chain, and the deficit equals the temporal variance of the complexity. The entropy is not a polynomial in the amplitudes (Krylov wavefunctions), and its late-time average is not determined by $p_n^\infty$ alone.

\subsection{Three levels with arbitrary weights}
\label{sec:N3psi}
 
\begin{figure}[t]
\hspace{1.5cm}
\includegraphics[width=0.87\textwidth]{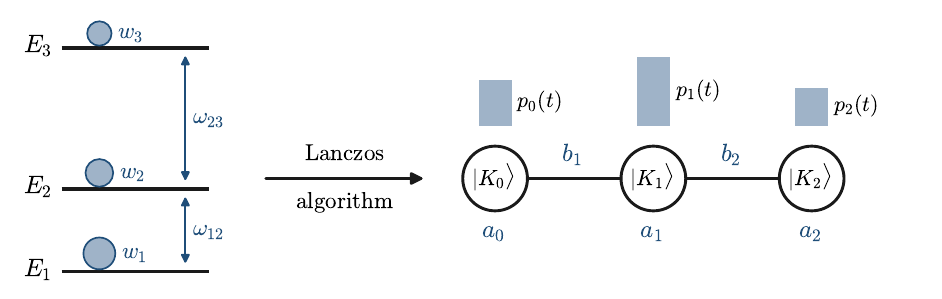}
\caption{Schematic of the Krylov construction for three levels. Left: energy levels $E_1<E_2<E_3$ with spacings $\omega_{12}$ and $\omega_{23}$, which define the ratio $r=\omega_{23}/\omega_{12}$. The disks indicate the spectral weights $w_k$ of the initial state. Right: the Lanczos algorithm maps the levels and their weights to a chain of three Krylov vectors with on-site coefficients $a_n$ and hopping amplitudes $b_n$. The bars show the Krylov distribution $p_n(t)$ at a fixed time $t$, whose mean is the spread complexity $K_S(t)$.}
\label{fig:schematic}
\end{figure}

For $d=3$, we order the energy levels as $E_1<E_2<E_3$ and write
\begin{align}
\omega_{ij}=E_j-E_i\,,\qquad \omega_{13}=\omega_{12}+\omega_{23}\,,\qquad \Pi=\omega_{12}\,\omega_{23}\,\omega_{13}\,.
\end{align}
With non-vanishing weights, the Krylov dimension is $\mathcal{K}=3$, and the Lanczos recursion terminates after two steps, so that the Krylov basis can be obtained directly (see Fig.\,\ref{fig:schematic} for a schematic representation). The initial vector has components $u_{0,k}=\sqrt{w_k}$, and the first Lanczos coefficient is the mean energy, $a_0=\bar E\equiv\sum_iw_iE_i$. As in the two-level case, the second vector follows from $b_1\ket{K_1}=(H-\bar E)\ket{K_0}$, whose norm fixes $b_1$. The third vector is orthogonal to the first two and is, up to a sign, their vector product in the energy basis. Thus, the Krylov basis reads
\begin{align}
u_{0,k}=\sqrt{w_k}\,,\qquad
u_{1,k}=\sqrt{\frac{w_k}{\Delta_2}}\,(E_k-\bar E)\,,\qquad
u_{2,k}=\sigma_k\sqrt{\frac{w_lw_m}{\Delta_2}}\;\omega_{lm}\,,
\label{eq:u2}
\end{align}
where $(k,l,m)$ is a permutation of $(1,2,3)$ with $l<m$, $\sigma_1=\sigma_3=1$, $\sigma_2=-1$, and
\begin{align}
\Delta_2=\sum_kw_k(E_k-\bar E)^2=\sum_{i<j}w_iw_j\,\omega_{ij}^2\,,
\end{align}
is the variance of the energy in the initial state. The remaining Lanczos coefficients are given by $a_n=\bra{K_n}H\ket{K_n}$ and $b_2=\bra{K_2}H\ket{K_1}$, which yields
\begin{align}
b_1^2=\Delta_2\,,\qquad b_2=\frac{\sqrt{w_1w_2w_3}}{\Delta_2}\,\Pi\,,\qquad
a_1=\bar E+\frac{1}{\Delta_2}\sum_kw_k(E_k-\bar E)^3\,,
\label{eq:aspread}
\end{align}
together with $a_2=\mathrm{Tr} \,H-a_0-a_1$, where we used the identities $\sum_k \sigma_k \omega_{lm} = \sum_k \sigma_k \omega_{lm} E_k = 0$ and $\sum_k \sigma_k \omega_{lm} E_k^2 = \Pi$. The coefficient $b_2$ vanishes if two levels coincide or if one weight vanishes. In that case, the Krylov space is two-dimensional, and the results of Sec.\,\ref{sec:N2} apply. 

For the coherent Gibbs state \eqref{eq:tfd}, the weights $w_k=e^{-\beta E_k}/Z$ depend on the inverse temperature, and Eq.\,\eqref{eq:aspread} gives the Lanczos coefficients as explicit functions of $\beta$. Since $\partial_\beta w_k=-(E_k-\bar E)\,w_k$, they satisfy the open Toda equations \cite{Dymarsky:2019elm,Angelinos:2025drf,Takahashi:2025iol}
\begin{align}
\partial_\beta a_n=-\big(b_{n+1}^2-b_n^2\big)\,,\qquad \partial_\beta b_n=-\frac{1}{2}\,b_n\big(a_n-a_{n-1}\big)\,,
\label{eq:toda}
\end{align}
with $b_0=b_3=0$. Equation \eqref{eq:aspread} is an explicit solution of the three-site open Toda lattice. The flow is isospectral, so that $a_0+a_1+a_2$, $\sum_na_n^2+2\sum_nb_n^2$ and the determinant of the Lanczos matrix are independent of $\beta$. Equation \eqref{eq:aspread} gives
\begin{align}
b_1^4\,b_2^2=w_1w_2w_3\,\Pi^2=\frac{e^{-\beta(E_1+E_2+E_3)}}{Z^3}\,\Pi^2\,,
\end{align}
which is the Hankel determinant of the moments of the weights and plays the role of a tau function of the lattice. At low temperature, the flow orders the spectrum along the chain, $a_0\to E_1$, $a_1\to E_2$ and $a_2\to E_3$, while the hopping amplitudes vanish exponentially,
\begin{align}
b_1\simeq\omega_{12}\,e^{-\beta\omega_{12}/2}\,,\qquad b_2\simeq\frac{\omega_{23}\,\omega_{13}}{\omega_{12}}\,e^{-\beta\omega_{23}/2}\,.
\end{align}
This is the Krylov-space counterpart of the reduction to a two-level problem at low temperature discussed in Sec.\,\ref{sec:beta}.
 
Substituting Eq.\,\eqref{eq:u2} into Eq.\,\eqref{eq:amp}, the Krylov wavefunctions $\phi_1$ and $\phi_2$ are sums of three exponentials whose coefficients sum to zero. Using $|\sum_k c_k e^{-i E_k t}|^2 = -4 \sum_{i<j} c_ic_j \sin^2(\omega_{ij} t/2)$ for $\sum_k c_k = 0$, the probabilities are given by
\begin{align}
p_1(t)=-\frac{4}{\Delta_2}\sum_{i<j}w_iw_j(E_i-\bar E)(E_j-\bar E)\sin^2\frac{\omega_{ij}t}{2}\,,~~
p_2 (t)=\frac{4w_1w_2w_3\Pi}{\Delta_2}\sum_{i<j}\frac{\sigma_{ij}}{\omega_{ij}}\sin^2\frac{\omega_{ij}t}{2}\,,
\label{eq:p12}
\end{align}
with $\sigma_{12}=\sigma_{23}=1$, $\sigma_{13}=-1$, and $p_0(t)=|\mathcal{R}(t)|^2=1-p_1(t)-p_2(t)$. One can check that at early times $p_1 (t) \simeq b_1^2t^2$ and $p_2 (t) \simeq\tfrac14b_1^2b_2^2t^4$, and that for $w_3\to0$, Eq.\,\eqref{eq:p12} reduces to the two-level result \eqref{eq:N2S}. Since Eq.\,\eqref{eq:p12} holds for arbitrary weights, it determines $K_S(t)=p_1(t)+2p_2(t)$ and $\sum_n n^2p_n(t)=p_1(t)+4p_2(t)$, and hence $V_S(t)$ and $E_S(t)$. In particular, the spread complexity is
\begin{align}
K_S(t)=\frac{4}{\Delta_2}\sum_{i<j}\Big[-w_iw_j(E_i-\bar E)(E_j-\bar E)+\frac{2\sigma_{ij}\,w_1w_2w_3\,\Pi}{\omega_{ij}}\Big]\sin^2\frac{\omega_{ij}t}{2}\,,
\label{eq:CSgen}
\end{align}
which is a superposition of three oscillations at the frequencies $\omega_{12}, \omega_{23}$ and $\omega_{13} = \omega_{12} +  \omega_{23}$, \emph{i.e.}, at all pairwise differences of the energy levels, while $V_S(t)$ additionally contains oscillations at their sums, differences, and doubled frequencies.

Since $E_S(t)$ is not a polynomial in the probabilities, we cannot obtain its late-time value from the identities \eqref{eq:timeavg} and must compute it separately. With $s_{n,k}=\sqrt{w_k}\,u_{n,k}$ and the overall phase $e^{-iE_1t}$ removed from Eq.\,\eqref{eq:amp}, each probability takes the form
\begin{align}
p_n(t)=\big|s_{n,1}+s_{n,2}\,e^{-i\chi}+s_{n,3}\,e^{-i\chi'}\big|^2\,,\qquad \chi=\omega_{12}t\,,\qquad \chi'=\omega_{13}t\,,
\label{eq:pnphase}
\end{align}
and depends on time only through the phases $\chi$ and $\chi'$, taken modulo $2\pi$. For an irrational ratio $\omega_{12}/\omega_{23}$, the point $(\chi,\chi')$ covers the square $[0,2\pi)^2$ uniformly as $t$ increases, and the late-time average of any function of the probabilities equals its average over two independent, uniformly distributed phases
\begin{align}
X^\infty=\frac{1}{4\pi^2}\int_0^{2\pi}\mathrm{d}\chi\int_0^{2\pi}\mathrm{d}\chi'\,X(\chi,\chi')\,.
\label{eq:torus}
\end{align}
Since $p_n(t)$ is invariant under an overall change of sign and the phase averages are invariant under shifts of $\chi$ and $\chi'$ by $\pi$, the signs of the $s_{n,k}$ can be absorbed into these phases, and only the moduli $a=|s_{n,1}|$, $b=|s_{n,2}|$ and $c=|s_{n,3}|$ enter.

The two averages in Eq.\,\eqref{eq:torus} can be performed successively. With $p_n=|\xi|^2$, $\xi=\zeta+c\,e^{-i\chi'}$ and $\zeta=a+b\,e^{-i\chi}$, the contribution of the $n$-th site to the entropy is $-p_n\ln p_n=-|\xi|^2\ln|\xi|^2$. For fixed $\zeta$, the average of $\ln|\zeta+c\,e^{-i\chi'}|$ over $\chi'$ equals $\ln\max(|\zeta|,c)$, which
follows by expanding $\ln |\xi|$ in powers of $c/|\zeta|$ for $|\zeta|>c$, and of $|\zeta|/c$ otherwise. One then obtains
\begin{align}
\frac{1}{4\pi}\int_0^{2\pi}\mathrm{d}\chi'\,|\xi|^2\ln|\xi|^2=F(|\zeta|,c)\,,~~ F(R,c)=(R^2+c^2)\ln\max(R,c)+\min(R,c)^2\,.
\label{eq:Favg}
\end{align}
The remaining integrand depends on $\chi$ only through $|\zeta|^2=a^2+b^2+2ab\cos\chi$, which is even in $\chi$, so that the late-time entropy reduces to a single integral,
\begin{align}
E_S^\infty=-2\sum_{n=0}^2M\big(|s_{n,1}|,|s_{n,2}|,|s_{n,3}|\big)\,,~~
M(a,b,c)=\frac1\pi\int_0^\pi\mathrm{d}\chi\,F\Big(\sqrt{a^2+b^2+2ab\cos\chi},\,c\Big)\,.
\label{eq:Hinfgen}
\end{align}
The function $M(a,b,c)$ is the phase average of $|a+b\,e^{-i\chi}+c\,e^{-i\chi'}|^2\ln|a+b\,e^{-i\chi}+c\,e^{-i\chi'}|$. Integrals of this type also appear in the theory of planar random walks and of Mahler measures \cite{Smyth_1981, Borwein2011}. Equation \eqref{eq:Hinfgen} is evaluated for three levels at infinite temperature in Sec.\,\ref{sec:entropy}.

\section{Three levels at infinite temperature}
\label{sec:shape}
 
\subsection{Scale and shape of the spectrum}
 
At $\beta=0$, the weights of the state \eqref{eq:tfd} are uniform, $w_i=\tfrac13$, and the Lanczos coefficients and amplitudes depend only on the energy differences $\omega_{ij}$. The centered energies $\varepsilon_i=E_i-\bar E$ (so that $\omega_{ij} = E_j - E_i = \varepsilon_j - \varepsilon_i)$ lie in the plane $\sum_i\varepsilon_i=0$, and we parametrize them as
\begin{align}
\varepsilon_k=\tfrac23\sqrt{Q}\,\cos\theta_k\,,\qquad \theta_k=\theta-\frac{2\pi k}{3}\,,\qquad k=1,2,3 \,.
\label{eq:param}
\end{align}
Here $Q=\tfrac32\sum_i\varepsilon_i^2=\tfrac12(\omega_{12}^2+\omega_{23}^2+\omega_{13}^2)$ is the scale of the spectrum, and $\theta$ is the polar angle in the plane of centered energies, which determines its shape. The ordering $E_1 <E_2 <E_3$
corresponds to $-\pi/3 < \theta < 0$.  Using trigonometric identities, we get
\begin{align}
\Pi^2=\frac{4}{27}\,Q^3\sin^23\theta\,,\qquad \varepsilon_1\varepsilon_2\varepsilon_3=\frac{2}{27}\,Q^{3/2}\cos3\theta \,.
\end{align}
It is convenient to use the shape variable $y=|\sin3\theta|\in[0,1]$, which vanishes when two levels are degenerate and equals one when the levels are equally spaced. In terms of the ratio of consecutive spacings $r=\omega_{23}/\omega_{12}$,
\begin{align}
y^2=\sin^2 3\theta = \frac{27\,r^2(1+r)^2}{4\,(1+r+r^2)^3} = \frac{27}{4\,Q_r^3} \,r^2(1+r)^2 \,,
\label{eq:yr}
\end{align}
where $Q_r=1+r+r^2$. This expression is invariant under $r\to1/r$, so $y$ is in one-to-one correspondence with $\min(r,1/r)$. It is easy to check that $r=0$ or $r\to\infty$ (\textit{i.e.}, $y=0$) is the degenerate limit and $r=1$ (\textit{i.e.}, $y=1$) is the equal-spacing limit.

The Lanczos coefficients follow from Eq.\,\eqref{eq:aspread}. Using the identities $\sum_k \varepsilon_k = 0$ and $\sum_k \varepsilon_k^3 =3\,\varepsilon_1\varepsilon_2\varepsilon_3$, one finds
\begin{align}
b_1^2=\tfrac29Q,\qquad b_2^2=\tfrac19y^2Q,\qquad a_0=\bar E,\qquad a_1-\bar E=-(a_2-\bar E)=\tfrac13\sqrt{Q}\,\cos3\theta \,.
\label{eq:bshapeS}
\end{align}
The coefficient $b_2$ vanishes for a degenerate pair, where the Krylov space becomes two-dimensional. The Krylov vectors follow from Eq.\,\eqref{eq:u2} with $w_k=\tfrac13$ and $\Delta_2=\tfrac29Q$. For $u_{1,k}$ this gives $u_{1,k}=\varepsilon_k/\sqrt{2Q/3}=\sqrt{2/3}\,\cos\theta_k$. For $u_{2,k}$, Eq.\,\eqref{eq:u2} gives $u_{2,k}=\sigma_k\,\omega_{lm}/\sqrt{2Q}$, and the energy differences follow from Eq.\,\eqref{eq:param} with $\cos A-\cos B=-2\sin\frac{A+B}{2}\sin\frac{A-B}{2}$. For instance, $\omega_{23}=\tfrac23\sqrt Q\,(\cos\theta_3-\cos\theta_2)=-\tfrac{2}{\sqrt3}\sqrt Q\,\sin\theta_1$, and in the same way $\sigma_k\omega_{lm}=-\tfrac{2}{\sqrt3}\sqrt Q\,\sin\theta_k$ for all $k$. Hence
\begin{align}
u_{0,k}=\frac{1}{\sqrt3},\qquad u_{1,k}=\sqrt{\frac23}\,\cos\theta_k,\qquad u_{2,k}=-\sqrt{\frac23}\,\sin\theta_k\,.
\label{eq:rigid}
\end{align}
The Krylov basis is therefore obtained from a fixed frame by a rotation through the angle $\theta$ in the plane orthogonal to $\ket{K_0}$. With Eq.\,\eqref{eq:amp}, the probabilities are
\begin{align}
p_0=\frac19\bigg|\sum_ke^{-i\varepsilon_kt}\bigg|^2,\qquad p_1=\frac29\bigg|\sum_k\cos\theta_k\,e^{-i\varepsilon_kt}\bigg|^2,\qquad p_2=\frac29\bigg|\sum_k\sin\theta_k\,e^{-i\varepsilon_kt}\bigg|^2 \,.
\label{eq:pbeta0}
\end{align}
Since $\varepsilon_k\propto\sqrt Q$, the scale sets the time $\sqrt Q\,t$, and the shape enters only through $\theta$. In terms of the ratio $r$ and the phase $\tau=\omega_{12}t$, with the levels at $0$, $1$ and $1+r$ in units of $\omega_{12}$ and $Q_r=1+r+r^2$, the same probabilities read (with $p_n$ now regarded as functions of $\tau$)
\begin{align}
\begin{aligned}
p_0&=\frac{1}{9}\,\bigg|1+e^{-i\tau}+e^{-i(1+r)\tau}\bigg|^2\,,\\
p_1&=\frac{1}{18\,Q_r}\,\bigg|(2+r)-(1-r)e^{-i\tau}-(1+2r)e^{-i(1+r)\tau}\bigg|^2\,,\\
p_2&=\frac{1}{6\,Q_r}\,\bigg|r-(1+r)e^{-i\tau}+e^{-i(1+r)\tau}\bigg|^2 \,.
\end{aligned}
\label{eq:pr}
\end{align}
The coefficients in $p_1$ and $p_2$ are those of $\phi_1$ and $\phi_2$ in Eq.\,\eqref{eq:u2}, expressed in units of $\omega_{12}$. In particular, the coefficients $(r,\,-(1+r),\,1)$ in $p_2$ are $\sigma_k\omega_{lm}$ for $k=1,2,3$. Their signs are fixed by the ordering of the levels, which restricts $\theta$ to $-\pi/3<\theta<0$, so that $\sin\theta_2>0$ while $\sin\theta_1,\sin\theta_3<0$. They sum to zero, as required by the orthogonality of $\ket{K_2}$ to $\ket{K_0}$.
 
\subsection{Exact spread complexity and variance}
 
The spread complexity and the second moment follow from Eq.\,\eqref{eq:pr}. We measure frequencies in units of $\omega_{12}$. With $r=\omega_{23}/\omega_{12}$, the three energy differences are $\omega_{12},~ \omega_{23}=r\,\omega_{12}$ and $\omega_{13}=\omega_{12}+\omega_{23}=(1+r)\,\omega_{12}$ so that they correspond to $f\in\{1,r,1+r\}$, and $Q=\omega_{12}^2Q_r$ with $Q_r=1+r+r^2$. Expanding the moduli in Eq.\,\eqref{eq:pr}, each probability is a constant plus three cosines at the frequencies $f\omega_{12}$. The term at frequency $f$ arises from the pair of levels separated by $f\omega_{12}$, with a coefficient given by the product of the corresponding two coefficients in Eq.\,\eqref{eq:pr}. Combining $p_1(t)+2p_2(t)$ and $p_1(t)+4p_2(t)$, we obtain
\begin{align}
K_S(t)&=1-\frac{1}{9Q_r} \sum_{f\in\{1,r,1+r\}} \alpha_f \cos(f\omega_{12}t)\,, \label{eq:KSn2}\\
\sum_n n^2p_n(t)&=\frac{5}{3}-\frac{1}{9Q_r}\sum_{f\in\{1,r,1+r\}} \mu_f \cos(f\omega_{12}t)\,, \label{eq:n2}
\end{align}
where the coefficients $\alpha_f$ and $\mu_f$ are given by
\begin{align}
\begin{aligned}
&\alpha_1=2+5r+5r^2\,,&\quad &\alpha_r=5+5r+2r^2\,,&\quad &\alpha_{1+r}=2-r+2r^2\,,\\
&\mu_1=2+11r+11r^2\,,&\quad &\mu_r=11+11r+2r^2\,,&\quad &\mu_{1+r}=2-7r+2r^2 \,. \label{alphamu}
\end{aligned}
\end{align}
The variance $V_S(t)=\sum_nn^2p_n(t)-K_S(t)^2$ follows from Eqs.\,\eqref{eq:KSn2} and \eqref{eq:n2}.\footnote{For equal weights, the spread complexity of three levels for a single spectrum was also obtained in Ref.\,\cite{Caputa:2025ozd} in the context of symmetry-resolved complexity. Eq.\,\eqref{eq:KSn2} agrees with Eq.\,(4.64) of Ref.\,\cite{Caputa:2025ozd}, with $D=2\omega_{12}^2Q_r$, $M_{12}=2\omega_{12}^2\alpha_1$, $M_{23}=2\omega_{12}^2\alpha_r$ and $M_{13}=2\omega_{12}^2\alpha_{1+r}$. Writing the result in terms of $r$ allows the ensemble average over the Gaussian ensembles in Sec.\,\ref{sec:rmt} to be performed.} Since $K_S(t)^2$ contains products of the cosines, $V_S(t)$ contains, in addition, the second harmonics and the sums and differences of the three frequencies,
\begin{align}
V_S(t)=\frac{25}{54}+\frac{1}{162\,Q_r^2}\Big[&\,2\sum_f\nu_f\cos(f\omega_{12}t)-\sum_f\alpha_f^2\cos(2f\omega_{12}t)-2\alpha_1\alpha_r\cos\big((1-r)\omega_{12}t\big)\nonumber\\
&-2\alpha_1\alpha_{1+r}\cos\big((2+r)\omega_{12}t\big)-2\alpha_r\alpha_{1+r}\cos\big((1+2r)\omega_{12}t\big)\Big]\,,
\label{eq:VS}
\end{align}
with the following values of $\nu_f$:
\begin{gather}
\nu_1 =8+4r-9r^2-26r^3-13r^4\,,\qquad \nu_r=-13-26r-9r^2+4r^3+8r^4\,,\nonumber\\
\nu_{1+r} =8+28r+27r^2+28r^3+8r^4\,.
\end{gather}
The reflection $E\to-E$ exchanges the two adjacent energy gaps and corresponds to $r\to1/r$. Under this transformation, $Q_r\to Q_r/r^2$ and $\omega_{12}\to r\,\omega_{12}$, while $\alpha_1\leftrightarrow\alpha_r$ and $\nu_1\leftrightarrow\nu_r$ up to the factors $r^{-2}$ and $r^{-4}$, respectively. Analogously, $\alpha_{1+r}$ and $\nu_{1+r}$ are multiplied by the same factors. Thus, the contributions associated with the two gaps are interchanged, and $K_S(t)$ and $V_S(t)$ remain invariant.

The late-time values follow from Eqs.\,\eqref{eq:KSn2} and \eqref{eq:VS}. For $\omega_{12}\neq\omega_{23}$, \emph{i.e.}, $r\neq1$, all frequencies in these expressions are nonzero, and the cosines average to zero at late times. The late-time values of $K_S$ and $V_S$ are given by the constant terms. The temporal variance of $K_S$ is the late-time average of the square of its oscillating part, \emph{i.e.}, one half of the sum of the squared Fourier coefficients in Eq.\,\eqref{eq:KSn2}. With $\sum_f\alpha_f^2=33\,Q_r^2$, we obtain
\begin{align}
K_S^\infty=1\,,\qquad \delta K_S^2=\frac{1}{2}\,\frac{1}{81\,Q_r^2}\sum_f\alpha_f^2=\frac{11}{54}\,,\qquad V_S^\infty=\frac{25}{54}\,,
\label{eq:univ}
\end{align}
for every spectrum with $r\neq1$. Thus, the late-time complexity and variance, as well as the temporal fluctuations of the complexity, are independent of the spectrum, and $V_S^{\infty}+\delta K_S^2=\frac{2}{3}$, in agreement with the sum rule Eq.\,\eqref{llt} for $d=3$.
 
At exactly equal spacing, $r=1$, the frequencies $\omega_{12}$ and $\omega_{23}$ coincide, and the same computation gives $\delta K_S^2=\frac{65}{162}$, $V_S^{\infty}=\frac{43}{162}$ and $V_S^{\infty}+\delta K_S^2=\frac{2}{3}$. This value is approached only on time scales of order $1/|\omega_{12}-\omega_{23}|$, since the frequency $|1-r|\,\omega_{12}$ in Eq.\,\eqref{eq:VS} vanishes as $r \to 1$.
 
\subsection{Late-time value of the Krylov entropy}
\label{sec:entropy}
 
We can compute the late-time value of the entropy from Eq.\,\eqref{eq:Hinfgen}. With $w_k=\frac{1}{3}$ and Eq.\,\eqref{eq:rigid}, the coefficients $s_{n,k}=\sqrt{w_k}\,u_{n,k}$ are
\begin{align}
s_{0,k}=\frac{1}{3}\,,\qquad s_{1,k}=\frac{\sqrt2}{3}\cos\theta_k\,,\qquad s_{2,k}=-\frac{\sqrt2}{3}\sin\theta_k\,.
\label{eq:sbeta0}
\end{align}
They do not depend on the scale $Q$, so that $E_S^\infty$ depends on the spectrum only through the angle $\theta$.
 
The contribution of $p_0$ is the same for every spectrum. Here $p_0=\frac{1}{9}\big|1+e^{-i\chi}+e^{-i\chi'}\big|^2$ is the squared distance from the origin after a planar walk of three unit steps in independent, uniformly distributed directions. In Eq.\,\eqref{eq:Hinfgen}, we set $a=b=c=\frac{1}{3}$, so that $R^2=a^2+b^2+2ab\cos\chi=\frac{4}{9}\cos^2\frac{\chi}{2}$, \emph{i.e.}, $R=\frac{2}{3}\cos\frac{\chi}{2}$ for $0\le\chi\le\pi$. Thus $R>c$ for $\chi<\frac{2\pi}{3}$ and $R<c$ for $\chi>\frac{2\pi}{3}$. With $R^2+c^2=\frac{1}{9}(3+2\cos\chi)$, the function $F$ of Eq.\,\eqref{eq:Favg} reads
\begin{align}
F(R,c)=-\frac{\ln3}{9}\,(3+2\cos\chi)+\frac{1}{9}\begin{cases}(3+2\cos\chi)\ln\big(2\cos\frac{\chi}{2}\big)+1\,, & 0\le\chi<\frac{2\pi}{3}\,,\\[1ex] 2+2\cos\chi\,, & \frac{2\pi}{3}<\chi\le\pi\,.\end{cases}
\end{align}
The contribution of $p_0$ is then given by
\begin{align}
    -\big(p_0\ln p_0\big)^{\infty} &= -\frac{2}{\pi} \int_{0}^{\pi} \mathrm{d}\chi\, F(R,c) = \frac{2 \ln 3}{9 \pi} \int_{0}^{\pi} \mathrm{d}\chi\, (3+2\cos \chi) \nonumber \\
    &- \frac{2}{9 \pi} \int_{0}^{2\pi/3} \mathrm{d}\chi\,\big[(3+2\cos\chi)\ln\big(2\cos\frac{\chi}{2}\big)+1\big] - \frac{2}{9 \pi} \int_{2\pi/3}^{\pi} \mathrm{d}\chi\,(2+2 \cos \chi)\,.
\end{align}
Evaluating the integrals, we find
\begin{align}
-\big(p_0\ln p_0\big)^{\infty}=\frac{2}{3}\ln3-\frac{4}{9}+\frac{1}{\sqrt{3}\pi}-\frac{2}{3\pi}\,\mathrm{Cl}_2\Big(\frac{\pi}{3}\Big)\,,
\label{eq:H0}
\end{align}
where we have used the result
\begin{align}
\int_0^{2\pi/3}\mathrm{d}\chi\,\ln\Big(2\cos\frac{\chi}{2}\Big)=\mathrm{Cl}_2\Big(\frac{\pi}{3}\Big)\,.
\end{align}
Here $\mathrm{Cl}_2$ is the Clausen function, defined as $\mathrm{Cl}_2(x)= - \int_0^{x}\mathrm{d}z\, \ln\big|2\sin\frac{z}{2}\big| = \sum_{k\ge1}\frac{\sin(kx)}{k^2}$.

The contributions of $p_1$ and $p_2$ depend on $\theta$ only through $|\cos\theta_k|$ and $|\sin\theta_k|$, since the late-time average involves only the moduli of the coefficients. Permutations and reflections of the levels shift $\theta$ by multiples of $\frac{2\pi}{3}$ or change its sign, and only permute these moduli. Under $\theta\to\theta+\frac{\pi}{2}$, one has $\cos\theta_k\to-\sin\theta_k$ and $\sin\theta_k\to\cos\theta_k$, so that the coefficients of $p_1$ and $p_2$ are exchanged and $E_S^\infty$ is again unchanged. Hence $E_S^\infty$ has period $\frac{\pi}{6}$ in $\theta$, and since $\sin3(\theta+\frac{\pi}{2})=-\cos3\theta$,
\begin{align}
E_S^\infty(y)=E_S^\infty\big(\sqrt{1-y^2}\big).
\label{eq:duality}
\end{align}
This duality relates nearly degenerate spectra, $y\to0$, to nearly equally spaced spectra, $y\to1$. Its fixed point $y=1/\sqrt2$ corresponds to $r=\frac{\sqrt3-1}{2}$.

\begin{figure}[t]
\centering
\includegraphics[width=0.5\textwidth]{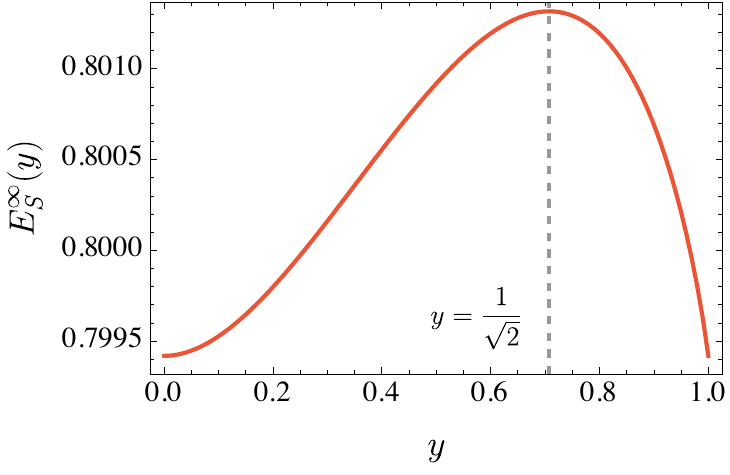}
\caption{Late-time value of the Krylov entropy for three levels at $\beta=0$ as a function of the shape variable $y$, Eq.\,\eqref{eq:Hinfgen}. The curve is symmetric under $y^2\to1-y^2$, Eq.\,\eqref{eq:duality}, and the dashed line marks the self-dual point $y=1/\sqrt2$. The end points are given by Eq.\,\eqref{eq:Hy0}.}
\label{fig:Sinf}
\end{figure}
 
The limit $y\to0$, \emph{i.e.}, $\theta\to0$, can be evaluated in closed form. In this limit, $\cos\theta_k\to\big(-\frac{1}{2},-\frac{1}{2},1\big)$ and $\sin\theta_k\to\big(-\frac{\sqrt3}{2},\frac{\sqrt3}{2},0\big)$. For $p_2$, only two coefficients remain, and $p_2=\frac{1}{6}|1-e^{-i\chi}|^2=\frac{2}{3}\sin^2\frac{\chi}{2}$. For $p_1$, we set $a=\frac{\sqrt2}{3}$ and $b=c=\frac{\sqrt2}{6}$ in Eq.\,\eqref{eq:Hinfgen}, so that $R^2=\frac{2}{9}\big(\frac{5}{4}+\cos\chi\big)\ge c^2$ and $F=(R^2+c^2)\ln R+c^2$ for all $\chi$. We find
\begin{align}
-\big(p_2\ln p_2\big)^{\infty}&=-\frac{1}{\pi}\int_0^{\pi}\mathrm{d}\chi\,\frac{2}{3}\sin^2\frac{\chi}{2}\,\ln\Big(\frac{2}{3}\sin^2\frac{\chi}{2}\Big)=\frac{1}{3}\big(\ln6-1\big)\,,\\
-\big(p_1\ln p_1\big)^{\infty}&=-\frac{2}{\pi}\int_0^{\pi}\mathrm{d}\chi\,F(R,c)=\frac{1}{3}\ln\frac{9}{2}-\frac{2}{9}\,.
\end{align}
Adding Eq.\,\eqref{eq:H0}, we obtain the closed-form expression
\begin{align}
E_S^{\infty}(0)=E_S^{\infty}(1)=\frac{5}{3}\ln3-1+\frac{1}{\sqrt{3}\pi}-\frac{2}{3\pi}\,\mathrm{Cl}_2\Big(\frac{\pi}{3}\Big)\approx0.7994\,,
\label{eq:Hy0}
\end{align}
where the value at $y=1$ follows from the duality \eqref{eq:duality}. Both values are understood as limits, since at $y=0$ two levels are degenerate and at $y=1$ the frequencies $\omega_{12}$ and $\omega_{23}$ coincide. More generally, for any rational $r$, the phases $(\chi,\chi')$ are confined to a closed curve on the torus, and the time average differs from Eq.\,\eqref{eq:Hinfgen}. Such spectra form a set of measure zero and do not affect the ensemble averages.
 
For other values of $y$, we evaluate Eq.\,\eqref{eq:Hinfgen} numerically. As shown in Fig.\,\ref{fig:Sinf}, $E_S^\infty(y)$ increases from the value \eqref{eq:Hy0} to a maximum of $E_S^\infty\approx0.8013$ at the self-dual point $y = \frac{1}{\sqrt{2}}$, and then decreases symmetrically in $y^2$. The maximum lies below $\ln3$, which is the entropy of the time-averaged distribution $p_n^{\infty}=1/3$.

\section{Gaussian random-matrix ensembles}
\label{sec:rmt}

In this section, we compute the ensemble averages of the spread complexity, variance, and entropy of the Krylov distribution for Gaussian random-matrix ensembles. In the remainder of the paper, $K_S, V_S$ and $E_S$ denote ensemble averages unless stated otherwise.
 
\subsection{Joint distribution of the energy levels}
\label{sec:jpdf}
 
We take the joint density of the three eigenvalues to be
\begin{align}
\mathcal{P}(E_1,E_2,E_3)\propto e^{-\frac{d \beta_D}{4}\sum_iE_i^2}\prod_{j<k}|E_j-E_k|^{\beta_D}\,,\qquad d=3\,,
\label{eq:jpdf}
\end{align}
with Dyson index $\beta_D=1,2,4$ for the Gaussian orthogonal, unitary and symplectic ensembles (GOE, GUE, GSE). As a reference without level repulsion, we also consider three independent standard normal levels, denoted IID, for which $\mathcal{P}\propto e^{-\sum_iE_i^2/2}$. This corresponds to $\beta_D=0$ with a different Gaussian width.
 
As shown in Sec.\,\ref{sec:shape}, at infinite temperature the Krylov-space quantities depend on the spectrum only through the spacing $\omega_{12}$ and the ratio $r=\omega_{23}/\omega_{12}$. We therefore change variables from $(E_1,E_2,E_3)$, with $E_1<E_2<E_3$, to $(\bar E,\omega_{12},r)$, where $E_2=E_1+\omega_{12}$ and $E_3=E_2+r\,\omega_{12}$. The volume element transforms as
\begin{align}
\mathrm{d}E_1\,\mathrm{d}E_2\,\mathrm{d}E_3 = \left|
\frac{\partial(E_1,E_2,E_3)}
{\partial(\bar E,\omega_{12},r)}
\right|
\mathrm{d}\bar E\,\mathrm{d}\omega_{12}\,\mathrm{d}r = \omega_{12} \,\mathrm{d}\bar E\,\mathrm{d}\omega_{12}\,\mathrm{d}r\,.
\end{align}
The repulsion factor becomes
\begin{align}
    \prod_{j<k}|E_j-E_k|^{\beta_D} = |E_1 - E_2|^{\beta_D} |E_1 - E_3|^{\beta_D} |E_2 - E_3|^{\beta_D} = \omega_{12}^{3 \beta_D} (r+r^2)^{\beta_D}\,,
\end{align}
and $\sum_iE_i^2=3\bar E^2+\frac{2}{3}\,Q_r\,\omega_{12}^2$ with $Q_r=1+r+r^2$. After integrating over the center of mass $\bar E$, we obtain
\begin{align}
\mathcal{P}(\omega_{12},r)\propto\omega_{12}^{3\beta_D+1}\,e^{-\kappa_{\beta_D}Q_r\,\omega_{12}^2}\,(r+r^2)^{\beta_D}\,,
\label{eq:density}
\end{align}
with $\kappa_{\beta_D}=\beta_D/2$ for the Gaussian ensembles and $\kappa_0=1/3$ for IID levels.
 
The distribution of the ratio follows by integrating Eq.\,\eqref{eq:density} over $\omega_{12}$. For this, we use
\begin{align}
\int_0^\infty\mathrm{d}\omega_{12}\,\omega_{12}^{2\gamma-1}e^{-a\,\omega_{12}^2}=\frac{\Gamma(\gamma)}{2\,a^{\gamma}}\,,\qquad \gamma,a>0\,,
\label{eq:gammaint}
\end{align}
with $a=\kappa_{\beta_D}Q_r$ and $\gamma=1+\frac{3\beta_D}{2}$, so that $2\gamma-1=3\beta_D+1$. This gives
\begin{align}
P_{\beta_D}(r)=\frac{1}{\mathcal{Z}_{\beta_D}}\,\frac{(r+r^2)^{\beta_D}}{(1+r+r^2)^{1+3\beta_D/2}}\,,
\label{eq:atas}
\end{align}
with the normalization
\begin{align}
\mathcal{Z}_{\beta_D}=\begin{cases}\frac{2\pi}{3\sqrt3}, & \beta_D=0\ \text{(IID)}\,,\\[1ex] \frac{8}{27}, & \beta_D=1\ \text{(GOE)}\,,\\[1ex] \frac{4\pi}{81\sqrt3}, & \beta_D=2\ \text{(GUE)}\,,\\[1ex] \frac{4\pi}{729\sqrt3}, & \beta_D=4\ \text{(GSE)}\,.\end{cases}
\label{eq:constants}
\end{align}
The width of the distribution of $\omega_{12}$ in Eq.\,\eqref{eq:density} depends on $r$, so the spacing and the ratio are not independent. The scale $Q=Q_r\,\omega_{12}^2$ of the spectrum, on the other hand, is independent of the ratio. To see this, we change variables from $\omega_{12}$ to $Q$ at fixed $r$, with $|\frac{\partial \omega_{12}}{\partial Q}| = \frac{1}{2 \sqrt{Q Q_r}}$, and obtain
\begin{align}
    \mathcal{P}(Q,r) = \bigg|\frac{\partial \omega_{12}}{\partial Q}\bigg| \,\mathcal{P}(\omega_{12},r) \propto
Q^{3\beta_D/2}e^{-\kappa_{\beta_D}Q}\,
\frac{(r+r^2)^{\beta_D}}
{(1+r+r^2)^{1+3\beta_D/2}}\,. \label{gdist}
\end{align}
Hence the joint distribution factorizes, and $Q$ and $r$ are statistically independent. The scale $Q$ follows a Gamma distribution with shape parameter $1+\frac{3 \beta_D}{2}$ and rate $\kappa_{\beta_D}$, and it sets the unit of time. For the Lanczos coefficients \eqref{eq:bshapeS}, this means that $b_1^2=\frac{2}{9}Q$ is independent of $b_2/b_1=y/\sqrt{2}$, which depends only on $r$ through Eq.\,\eqref{eq:yr}.

The factorization also gives the statistics of the Lanczos coefficients for the Gaussian ensembles. In the parametrization \eqref{eq:param}, the density \eqref{eq:jpdf} depends on the angle $\theta$ only through the factor $|\sin3\theta|^{\beta_D}$, which gives $\langle\cos^23\theta\rangle=1/(\beta_D+2)$. The reflection $E\to-E$ changes the sign of $\cos3\theta$, so all $a_n$ have zero mean. Since $\bar E$, $Q$ and $\theta$ are independent, Eq.\,\eqref{eq:bshapeS} gives
\begin{align}
\mathrm{Var}(a_1)=\mathrm{Var}(a_2)=\frac{5\beta_D+6}{9\beta_D(\beta_D+2)}\,,~~
\mathrm{Var}(b_1)=\frac{4}{9\beta_D}\bigg[1+\frac{3\beta_D}{2}-\frac{\Gamma\big(\frac32+\frac{3\beta_D}{2}\big)^2}{\Gamma\big(1+\frac{3\beta_D}{2}\big)^2}\bigg]\,.
\label{eq:varab}
\end{align}
Writing $\mathrm{Var}(b_1)=c_{\beta_D}/(9\beta_D)$, where $c_{\beta_D}=0.95$, $0.97$ and $0.98$ for the GOE, the GUE and the GSE, the ratio takes the form\footnote{We thank Javier Magan for suggesting this computation.}
\begin{align}
\frac{\mathrm{Var}(a_1)}{\mathrm{Var}(b_1)}=\frac{1}{c_{\beta_D}}\bigg(4+\frac{\beta_D-2}{\beta_D+2}\bigg)\,.
\label{eq:ratioab}
\end{align}
For large random matrices, the variance of the diagonal Lanczos coefficients is four times that of the off-diagonal ones \cite{Balasubramanian:2022dnj}. For three levels, this ratio is close to $4$.
 
For $\beta_D=1,2,4$, Eq.\,\eqref{eq:atas} is the distribution of the ratio of consecutive level spacings obtained in Ref.\,\cite{Atas2013distribution} from $3\times3$ Gaussian matrices; see also Ref.\,\cite{Oganesyan:2007wpd}. For $\beta_D=0$ it reduces to $P_0(r)=\mathcal{Z}_0^{-1}(1+r+r^2)^{-1}$, the ratio distribution of three independent Gaussian levels. This is different from the Poissonian result $(1+r)^{-2}$ \cite{Atas2013distribution}, which applies to levels with independent, exponentially distributed spacings. For three levels at infinite temperature, the spacing-ratio distribution plays the role that the Wigner surmise plays for two levels.
 
\subsection{Ensemble averages of the Krylov-space quantities} \label{avgg}
 
The spread complexity \eqref{eq:KSn2} depends on $\omega_{12}$ only through the cosines $\cos(f\omega_{12}t)$, while their coefficients $\frac{\alpha_f}{9Q_r}$ depend only on $r$. The ensemble average can therefore be taken in two steps: first over $\omega_{12}$ at fixed $r$, with the weight \eqref{eq:density}, and then over $r$, with the distribution \eqref{eq:atas}. For the first step, we expand the cosine in powers of $\omega_{12}$ and apply Eq.\,\eqref{eq:gammaint}, with $\gamma = 1 + \frac{3 \beta_D}{2} + m$, to each power $\omega_{12}^{2m}$. This gives
\begin{align}
\frac{\int_0^\infty\mathrm{d}\omega_{12}\,\omega_{12}^{3 \beta_D+1}e^{-\kappa_{\beta_D}Q_r\omega_{12}^2}\cos(f\omega_{12}t)}{\int_0^\infty\mathrm{d}\omega_{12}\,\omega_{12}^{3 \beta_D+1}e^{-\kappa_{\beta_D}Q_r\omega_{12}^2}}
&=\sum_{m=0}^\infty\frac{(-1)^m(ft)^{2m}}{(2m)!}\,\frac{\Gamma(1 + \frac{3 \beta_D}{2}+m)}{\Gamma(1 + \frac{3 \beta_D}{2})\,(\kappa_{\beta_D}Q_r)^m}\nonumber\\
&={}_1F_1\Big(1 + \frac{3 \beta_D}{2};\frac{1}{2};-\frac{f^2t^2}{4\kappa_{\beta_D}Q_r}\Big)\,,
\label{eq:cosavg}
\end{align}
where we used $(2m)!=4^m\,m!\,(\frac{1}{2})_m$, and ${}_1F_1(a;b;z)=\sum_m\frac{(a)_m}{(b)_m}\frac{z^m}{m!}$ is the confluent hypergeometric function. Thus, the average over the spacing replaces each cosine in Eq.\,\eqref{eq:KSn2} by a confluent hypergeometric function, and the average over the ratio gives
\begin{align}
K_S(t)=1-\frac{1}{9}\int_0^\infty\mathrm{d} r\,\frac{P_{\beta_D}(r)}{Q_r}\sum_f\alpha_f \,{}_1F_1\Big(1 + \frac{3 \beta_D}{2};\frac{1}{2};-\frac{f^2t^2}{4\kappa_{\beta_D}Q_r}\Big)\,.
\end{align}
The contribution of $r>1$ equals that of $r<1$. The reflection $E\to-E$ maps a spectrum with spacings $(\omega_{12},\omega_{23})$ to one with spacings $(\omega_{23},\omega_{12})$, so that $r\to1/r$. It leaves the eigenvalue density \eqref{eq:jpdf} invariant, and it does not change the probabilities $p_n(t)$, since the amplitudes $\phi_n(t)$ change at most by a sign and complex conjugation. Hence, we can write
\begin{align}
K_S(t)=1-\frac{2}{9}\int_0^1\mathrm{d} r\,\frac{P_{\beta_D}(r)}{Q_r}\sum_f\alpha_f \,{}_1F_1\Big(1 + \frac{3 \beta_D}{2};\frac{1}{2};-\frac{f^2t^2}{4\kappa_{\beta_D}Q_r}\Big)\,,
\label{eq:KSavg}
\end{align}
where the sum runs over $f\in\{1,r,1+r\}$. This is Eq.\,\eqref{eq:main}. The integration variable is now the ratio $\min(r,1/r)$ used in Ref.\,\cite{Atas2013distribution}, whose distribution on $(0,1)$ is $2P_{\beta_D}(r)$. Restricting the integral to $(0,1)$ also removes the slowly decaying tail of $P_{\beta_D}(r)$ at large $r$, which for $\beta_D=0$ decays only as $r^{-2}$.

At $\beta=0$, $p_0(t)=|\mathcal{R}(t)|^2$ is the normalized SFF. Since $K_S= p_1 + 2p_2 = 1-p_0+p_2$ for three levels (see Eq.\,\eqref{eq:pr}), the complexity contains, in addition to $p_0$, the population $p_2$ of the last Krylov site. The ensemble average of $p_0$ depends only on the distributions of the individual energy differences, whereas the coefficients of $p_2$ depend on the ratio $r$. Thus its average involves the joint distribution of the two spacings.

The variance is averaged in the same way as the complexity. Each cosine $\cos(x\,\omega_{12}t)$ in Eq.\,\eqref{eq:VS}, where $x$ is one of the frequencies in units of $\omega_{12}$, is replaced according to Eq.\,\eqref{eq:cosavg} by
\begin{align}
\mathcal{F}(x)={}_1F_1\Big(1+\frac{3\beta_D}{2};\frac{1}{2};-\frac{x^2t^2}{4\kappa_{\beta_D}Q_r}\Big)\,,\nonumber
\end{align}
while the constant $\frac{25}{54}$ is unchanged. This gives
\begin{align}
V_S(t)=\frac{25}{54}+\frac{1}{81}\int_0^1\mathrm{d} r\,\frac{P_{\beta_D}(r)}{Q_r^2}\Big[&\,2\sum_f\nu_f\,\mathcal{F}(f)-\sum_f\alpha_f^2\,\mathcal{F}(2f)-2\alpha_1\alpha_r\,\mathcal{F}(1-r)\nonumber\\
&-2\alpha_1\alpha_{1+r}\,\mathcal{F}(2+r)-2\alpha_r\alpha_{1+r}\,\mathcal{F}(1+2r)\Big]\,,
\label{eq:VSavg}
\end{align}
where both sums run over $f\in\{1,r,1+r\}$. The ensemble enters Eqs.\,\eqref{eq:KSavg} and \eqref{eq:VSavg} only through $\beta_D$, $P_{\beta_D}(r)$ and $\kappa_{\beta_D}$.
 
For the GOE, ${}_1F_1(\frac{5}{2};\frac{1}{2};-z)=e^{-z}(1-4z+\frac{4}{3}z^2)$, so the integrand is Gaussian in $t$. For IID, GUE and GSE, the first argument is an integer, and ${}_1F_1$ reduces to a polynomial plus a polynomial times $\sqrt z\,D_{+}(\sqrt z)$, where $D_{+}(x)=e^{-x^2}\int_0^xe^{u^2}\mathrm{d} u$ is the Dawson function. For instance, ${}_1F_1(1;\frac{1}{2};-z)=1-2\sqrt z\,D_{+}(\sqrt z)$. For two levels, the situation is reversed, \emph{i.e.}, the GOE average involves the Dawson function, and the GUE and GSE averages are Gaussian \cite{Caputa:2024vrn}. 

\begin{figure}[t]
\centering
\includegraphics[width=\textwidth]{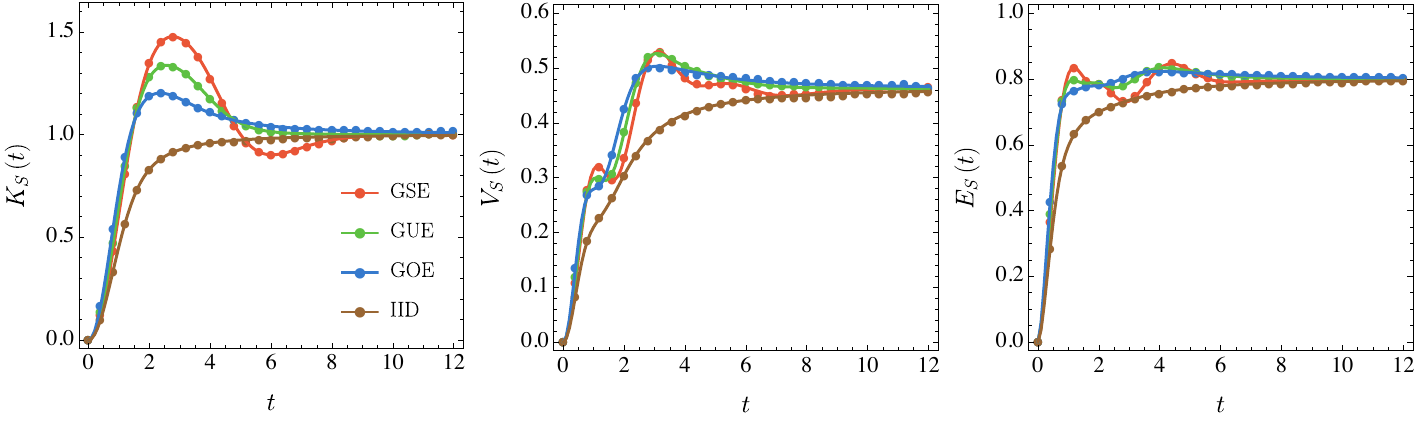}
\caption{Ensemble averages of (left) the spread complexity, (middle) the variance, and (right) the Krylov entropy at $\beta=0$ for three levels. The complexity and variance curves in the left and middle panels are computed from Eqs.\,\eqref{eq:KSavg} and \eqref{eq:VSavg}, and the entropy result in the right panel from Eq.\,\eqref{eq:SKavg} by quadrature. The circles are numerical results, obtained from spectra sampled with the tridiagonal $\beta$-Hermite ensembles \cite{Dumitriu:2002beta, Balasubramanian:2022dnj} and the uniform weights $w_k=1/3$. The numerical results are averages over $10^4$ realizations of the Hamiltonian.}
\label{fig:beta0}
\end{figure}
 
We compare Eqs.\,\eqref{eq:KSavg} and \eqref{eq:VSavg} with numerical results and find excellent agreement as shown in Fig.\,\ref{fig:beta0}. In these simulations, the energy levels were sampled as the eigenvalues of the tridiagonal $\beta$-Hermite ensembles \cite{Dumitriu:2002beta, Balasubramanian:2022dnj}, whose joint eigenvalue density coincides with Eq.\,\eqref{eq:jpdf} after appropriate rescaling of the eigenvalues. The Krylov basis was then constructed for the state with uniform weights $w_k=1/3$.\footnote{The off-diagonal elements of the tridiagonal matrix are the Lanczos coefficients of the initial vector $(1,0,0)$, whose spectral weights are random, and not those of the infinite-temperature state.}
 
The entropy is not a finite Fourier series, and the average over $\omega_{12}$ cannot be carried out in closed form. Its ensemble average is a two-dimensional integral over $\omega_{12}$ and $r$ with the weight \eqref{eq:density}. Normalizing the integral over $\omega_{12}$ with Eq.\,\eqref{eq:gammaint}, and restricting the integral over $r$ to $(0,1)$ by the same reflection argument, we obtain
\begin{align}
E_S(t)=-2\int_0^1\mathrm{d} r\,P_{\beta_D}(r)\int_0^\infty\mathrm{d}\omega_{12}\,\frac{2(\kappa_{\beta_D}Q_r)^{1+3\beta_D/2}}{\Gamma\big(1+\frac{3\beta_D}{2}\big)}\,\omega_{12}^{3\beta_D+1}e^{-\kappa_{\beta_D}Q_r\omega_{12}^2}\sum_np_n\ln p_n\,,
\label{eq:SKavg}
\end{align}
where the probabilities $p_n$ are given by Eq.\,\eqref{eq:pr} with $\tau=\omega_{12}t$. We evaluate this integral by quadrature.

\subsection{Time dependence of the ensemble averages}
 
\begin{figure}[t]
\centering
\includegraphics[width=0.48\textwidth]{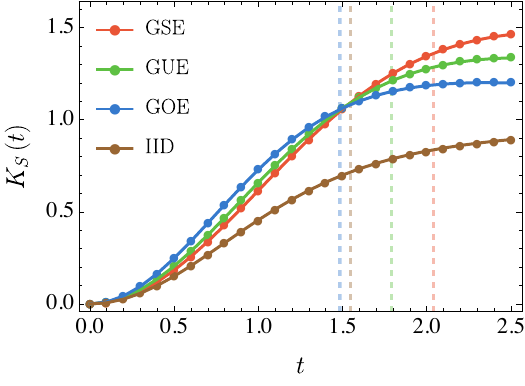}
\caption{Early-time behavior of the ensemble-averaged spread complexity at $\beta=0$ for three levels. The circles are numerical results as in Fig.\,\ref{fig:beta0} and the solid lines are from Eq.\,\eqref{eq:KSavg}. The dashed lines indicate the time $t_{*}$ of Eq.\,\eqref{tstar} for each ensemble. The numerical results are averages over $10^4$ realizations of the Hamiltonian.}
\label{fig:beta0earlyt}
\end{figure}

Figure \ref{fig:beta0} shows the ensemble averages of the spread complexity, the variance and the entropy. For a single spectrum, we have $K_S\simeq V_S\simeq b_1^2t^2$ at early times with $b_1^2=\frac{2}{9}Q$, so that the growth saturates the speed limit $\big|\frac{\mathrm{d}K_S}{\mathrm{d}t}\big|\le2b_1\sqrt{V_S}$ at leading order in $t$, as for any initial state. This bound is the analogue for states of the bound derived for operators in Ref.\,\cite{Hornedal:2022pkc}. To obtain the early-time behavior of the
complexity, we use the expansion of the hypergeometric series
\begin{align}{}_1F_1\Big(\gamma_{\beta_D};\frac{1}{2};-\frac{f^2t^2}{4\kappa_{\beta_D}Q_r}\Big) = 1 - \frac{\gamma_{\beta_D}}{2\kappa_{\beta_D} Q_r} f^2 t^2 + \frac{\gamma_{\beta_D}(\gamma_{\beta_D}+1)}{24 \kappa_{\beta_D}^2 Q_r^2} f^4 t^4 + O(t^6)\,,
\end{align}
where $\gamma_{\beta_D} = 1+3 \beta_D/2$. Denoting the sum $S_n = \sum_f \alpha_f f^n$, we can write
\begin{align}
 \sum_f \alpha_f \,{}_1F_1\Big(\gamma_{\beta_D};\frac{1}{2};-\frac{f^2t^2}{4\kappa_{\beta_D}Q_r}\Big) = S_0 - \frac{\gamma_{\beta_D}}{2\kappa_{\beta_D} Q_r} S_2 t^2 + \frac{\gamma_{\beta_D}(\gamma_{\beta_D}+1)}{24 \kappa_{\beta_D}^2 Q_r^2} S_4 t^4 + O(t^6)\,.
\end{align}
Using Eq.\,\eqref{alphamu}, we find
\begin{align}
    S_0 = \sum_f \alpha_f = 9Q_r\,,~~~  S_2 = \sum_f \alpha_f f^2 = 4 Q_r^2\,,~~~  S_4 = \sum_f \alpha_f f^4 = 4Q_r^3\Big(1-\frac{y^2}{3}\Big)\,,
\end{align}
where $y$ is given by Eq.\,\eqref{eq:yr}. Dividing by $Q_r$, the ensemble-averaged complexity in Eq.\,\eqref{eq:KSavg} becomes
\begin{align}
    K_S(t) = 1 - \frac{2}{9} \int_0^1 \mathrm{d}r P_{\beta_D}(r) \bigg[9 - \frac{2\gamma_{\beta_D}}{\kappa_{\beta_D}} t^2 + \frac{\gamma_{\beta_D}(\gamma_{\beta_D}+1)}{6 \kappa_{\beta_D}^2}\bigg(1-\frac{y^2}{3}\bigg) t^4 + O(t^6)\bigg]\,.
\end{align}
Using $\int_0^1P_{\beta_D}\,\mathrm{d}r=\frac{1}{2}$ and $2\int_0^1P_{\beta_D}\,y^2\,\mathrm{d}r=\frac{1+\beta_D}{2+\beta_D}$, we obtain
\begin{align}
K_S(t)=\frac{2+3\beta_D}{9\kappa_{\beta_D}}\,t^2-\frac{(2+3\beta_D)(4+3\beta_D)(5+2\beta_D)}{648\,\kappa_{\beta_D}^2\,(2+\beta_D)}\,t^4+O(t^6)\,.
\label{eq:short4}
\end{align}
The early-time growth of the complexity is shown in Fig.\,\ref{fig:beta0earlyt}. The quadratic growth thus holds for $t\ll t_*$, where $t_{*}$ is the time at which the $O(t^4)$ term becomes equal (in magnitude) to the $O(t^2)$ term. This gives
\begin{align}
t_{*}=\sqrt{\frac{72\,\kappa_{\beta_D}(2+\beta_D)}{(4+3\beta_D)(5+2\beta_D)}}\,, \label{tstar}
\end{align}
\emph{i.e.}, $t_*\approx1.55$, $1.48$, $1.79$ and $2.04$ for IID levels, the GOE, the GUE and the GSE. They are shown by the dashed lines in Fig.\,\ref{fig:beta0earlyt}. These values depend on the normalization of the eigenvalue density \eqref{eq:jpdf}.

Returning to Fig.\,\ref{fig:beta0}, for the Gaussian ensembles, all three (complexity, variance and entropy) averages rise above their late-time values before relaxing to the plateau, whereas for IID levels they approach their late-time values from below. The maximum of the variance is reached later than that of the complexity. In all three cases, the height of the maximum relative to the late-time value increases with $\beta_D$, and it decreases from $K_S$ to $V_S$ to $E_S$. The variance shows, in addition, a ``shoulder'' at early times, where the population of $|K_1\rangle$ is close to its maximum and that of $|K_2\rangle$ is still small. For two levels, the relative maxima of the averaged complexity are larger in each ensemble, and since $V_S(t)=K_S(2t)/4$ at $\beta=0$, the same holds for the variance. The overshoot is therefore weaker for three levels than for two levels \cite{Nandy:2024mml}. Since $P_{\beta_D}(r) \propto r^{\beta_D}$ at small $r$, part of the dependence of the peak on $\beta_D$ reflects the repulsion between adjacent levels, while the correlation between adjacent spacings enters through the full form of $P_{\beta_D}(r)$.

\subsection{Finite temperature and the low-temperature limit}
\label{sec:beta}
 
\begin{figure}[t]
\centering
\includegraphics[width=0.85\textwidth]{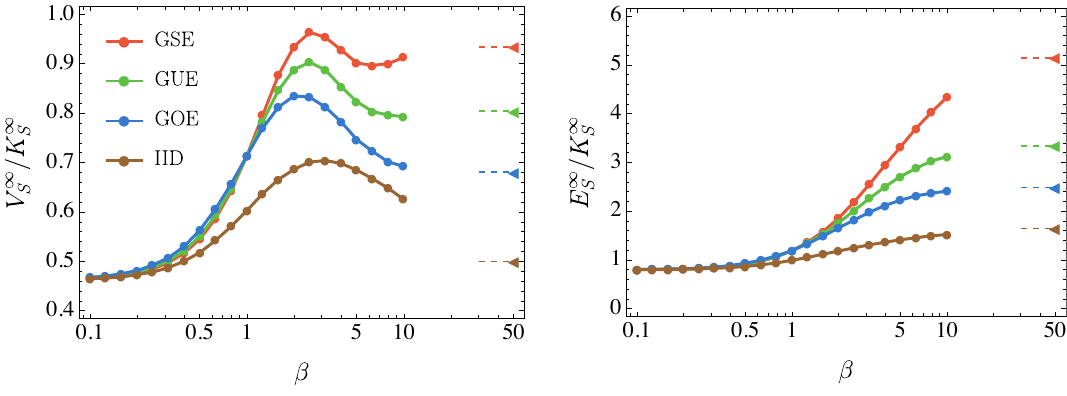}
\caption{Ratios of the ensemble-averaged late-time values of (left) the variance and the complexity and (right) the entropy and the complexity for three levels as functions of $\beta$. The numerical results are averages over $10^5$ realizations of the Hamiltonian and are restricted to $\beta\leq10$. The dashed lines and the triangles at the right edge mark the low-temperature limits $\beta\to\infty$, Eq.\,\eqref{eq:lowTV}.}
\label{fig:beta}
\end{figure}

For $\beta>0$, the weights of the state \eqref{eq:tfd} depend on the energies, and the factorization into scale and shape no longer applies. For each spectrum, the late-time values are given exactly by Eqs.\,\eqref{eq:timeavg} and \eqref{eq:Hinfgen}, and their ensemble averages were evaluated
numerically. The late-time value of the complexity decreases monotonically with $\beta$, since the state becomes concentrated on the ground level. The late-time values of the variance and of the entropy first increase, reach a maximum at $\beta$ of order unity, and then decrease. Figure \ref{fig:beta} shows the ratios of the three late-time values. The ratio of the variance to the complexity varies non-monotonically. It increases from $\frac{25}{54}$ at $\beta=0$ to a maximum at $\beta$ of order unity, whose height increases with $\beta_D$, and for the GUE and the GSE it then passes through a shallow minimum before it approaches its low-temperature limit. The ratio of the entropy to the complexity increases monotonically.
 
At low temperature, the complexities are dominated by spectra in which the two lowest levels are nearly degenerate, $\omega_{12}\lesssim1/\beta$. The weight of the third level is suppressed by $e^{-\beta\omega_{13}}$, and the problem reduces to the two-level problem of Sec.\,\ref{sec:N2}. With $u=\beta\omega_{12}$, the weights of the two lowest levels give
\begin{align}
w_1 = \frac{1}{1+e^{-u}}\,,~~ w_2 = \frac{e^{-u}}{1+e^{-u}}\,,~~ \Rightarrow ~~q\equiv w_1w_2=\frac{e^{-u}}{(1+e^{-u})^2}=\frac{1}{4}\,\mathrm{sech}^2\frac{u}{2}\,.
\end{align}
By Eq.\,\eqref{prob}, the probability on the second site is $P(t)=4q\sin^2x$ with $x=\frac{\omega_{12}t}{2}$. Since, by \eqref{eq:N2S}, the complexity and variance are $K_S(t) = P(t)$ and $V_S(t) = P(t) (1-P(t))$ respectively, the late-time values follow by averaging over $t$ (equivalently over $x$). For a single spectrum, $(\sin^2x)^{\infty}=\frac{1}{2}$ and $(\sin^4x)^{\infty}=\frac{3}{8}$ give
\begin{align}
K_S^{\infty}(u)=2q\,,~~~ V_S^{\infty}(u)=2q-6q^2\,,
\label{eq:KV2level}
\end{align}
where the argument $u$ indicates the values for a single spectrum. Similarly, the late-time value of the entropy is
\begin{align}
    E_S^{\infty} &= -\big(P(t)\ln P(t)\big)^{\infty} - \big[(1-P(t))\ln (1-P(t))\big]^{\infty}
    \nonumber \\
    &= -2q\,(\ln q+1) - 1+\sqrt{1-4q}+2q -2(1-2q)\ln\left(\frac{1+\sqrt{1-4q}}{2}
\right) \nonumber \\
    &= 2\ln\big(1+e^{-u}\big)+\frac{u}{2}\,\mathrm{sech}^2\frac{u}{2}-\frac{2}{1+e^{u}}\,. \label{eq:s2level}
\end{align}
At $u=0$, Eqs.\,\eqref{eq:KV2level} and \eqref{eq:s2level} reproduce the values $\frac{1}{2}$, $\frac{1}{8}$ and $2\ln2-1$ of Sec.\,\ref{sec:N2}.

For the ensemble averages, we need the distribution of the small spacing. As $\omega_{12}\to0$, the repulsion factor in Eq.\,\eqref{eq:jpdf} vanishes as $\omega_{12}^{\beta_D}$, and the other factors remain finite, so that the density of $\omega_{12}$ behaves as $C\,\omega_{12}^{\beta_D}$ with a constant $C$ that depends on the ensemble. Since the late-time values \eqref{eq:KV2level} and \eqref{eq:s2level} decay exponentially for $u\gg1$, only $\omega_{12}\lesssim1/\beta$ contributes for $\beta\to\infty$. Therefore,
\begin{align}
K_S^{\infty} \simeq
\int_0^{\infty} \mathrm{d} \omega_{12}\,
\mathcal{P}(\omega_{12})
K_S^{\infty}(\beta\omega_{12}) \simeq C\int_0^\infty \mathrm{d}\omega_{12}\,
\omega_{12}^{\beta_D}
K_S^\infty(\beta\omega_{12})\,.
\end{align}
Changing the variable $\beta\omega_{12} = u$ gives
\begin{align}
K_S^\infty\simeq\frac{C}{\beta^{\beta_D+1}}\int_0^\infty\mathrm{d}u\,u^{\beta_D}K_S^\infty(u)\,,
\end{align}
and analogously for $V_S^\infty$ and $E_S^\infty$. All three late-time values therefore decrease as $\beta^{-(\beta_D+1)}$, and the constant $C$ cancels in their ratios.
 
We expand in powers of $e^{-u}$, for instance
\begin{align}
q=\sum_{n\ge1}(-1)^{n+1}n\,e^{-nu}\,,\qquad q^2=\frac{1}{6}\sum_{n\ge1}(-1)^n\big(n^3-n\big)\,e^{-nu}\,,
\end{align}
and integrate term by term with $\int_0^\infty\mathrm{d}u\,u^me^{-nu}=\frac{m!}{n^{m+1}}$. Each term then gives a Dirichlet eta function $\eta(s)=\sum_{n\ge1}(-1)^{n+1}n^{-s}$, and we obtain
\begin{align}
\int_0^\infty\mathrm{d}u\,u^{\beta_D}q&=\beta_D!\,\eta(\beta_D)\,,\qquad
\int_0^\infty\mathrm{d}u\,u^{\beta_D}q^2=\frac{\beta_D!}{6}\big[\eta(\beta_D)-\eta(\beta_D-2)\big]\,,\nonumber\\
\int_0^\infty\mathrm{d}u\,u^{\beta_D}E_S^\infty(u)&=2\,\beta_D!\,\big[\eta(\beta_D+2)+\beta_D\,\eta(\beta_D+1)\big]\,.
\label{eq:etaint}
\end{align}
For $s\le0$, $\eta(s)$ is understood as its analytic continuation, $\eta(0)=\frac{1}{2}$, $\eta(-1)=\frac{1}{4}$ and $\eta(-2)=0$. Using Eqs.\,\eqref{eq:KV2level} and \eqref{eq:etaint}, the ratios of the late-time values approach
\begin{align}
\rho_V\equiv\lim_{\beta\to\infty}\frac{V_S^{\infty}}{K_S^{\infty}}=\frac{1}{2}+\frac{\eta(\beta_D-2)}{2\,\eta(\beta_D)}\,,~~~
\rho_E\equiv\lim_{\beta\to\infty}\frac{E_S^{\infty}}{K_S^{\infty}}=\frac{\eta(\beta_D+2)+\beta_D\,\eta(\beta_D+1)}{\eta(\beta_D)}\,.
\label{eq:lowT0}
\end{align}
With $\eta(1)=\ln2$, $\eta(3)=\frac{3}{4}\zeta(3)$, $\eta(5)=\frac{15}{16}\zeta(5)$ and $\eta(2k)=(1-2^{1-2k})\zeta(2k)$, one obtains for the four ensembles
\begin{align}
\begin{alignedat}{3}
&\beta_D=0:\qquad &\rho_V&=\frac{1}{2}\,, &\qquad \rho_E&=\frac{\pi^2}{6}\,,\\
&\beta_D=1: &\rho_V&=\frac{1}{2}+\frac{1}{8\ln2}\,, & \rho_E&=\frac{\pi^2+9\,\zeta(3)}{12\ln2}\,,\\
&\beta_D=2: &\rho_V&=\frac{1}{2}+\frac{3}{\pi^2}\,, & \rho_E&=\frac{7\pi^2}{60}+\frac{18\,\zeta(3)}{\pi^2}\,,\\
&\beta_D=4: &\rho_V&=\frac{1}{2}+\frac{30}{7\pi^2}\,, & \rho_E&=\frac{31\pi^2}{294}+\frac{2700\,\zeta(5)}{7\pi^4}\,.
\end{alignedat}
\label{eq:lowTV}
\end{align} Since only two levels contribute, these limits coincide with those of the two-level problem, and they hold for any $d$. In contrast to infinite temperature, the low-temperature ratios differ between the ensembles, because they are determined by the distribution of the smallest spacing, which is controlled by level repulsion.

\subsection{Approach to the late-time values}
\label{sec:late}

\begin{figure}[t]
\centering
\includegraphics[width=0.9\textwidth]{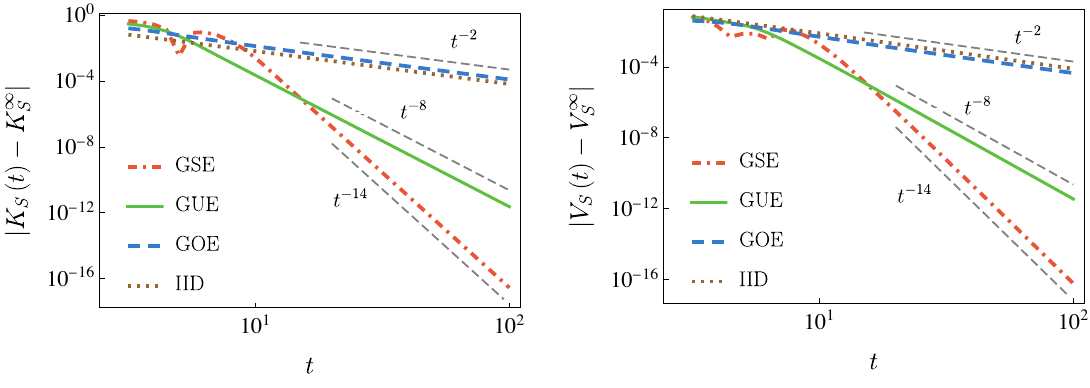}
\caption{Deviation of the ensemble averages of (left) the complexity and (right) the variance from their late-time values at $\beta=0$, on logarithmic axes. The dashed grey lines indicate the power laws \eqref{eq:late} as a guide to the eye.}
\label{fig:late}
\end{figure}

Figure \ref{fig:late} shows how the ensemble averages approach their late-time values at $\beta=0$. For a single spectrum, $K_S(t)$ and $V_S(t)$ oscillate indefinitely, and the averages relax only through the dephasing of different spectra. This dephasing is slowest for spectra with nearly degenerate levels, so that the decay is set by the eigenvalue density \eqref{eq:jpdf} close to degenerate configurations. Near a degenerate pair, the density is proportional to $|\omega_{23}|^{\beta_D}$, which is not smooth for odd $\beta_D$ and leads to a decay $t^{-(\beta_D+1)}$. Near the triple degeneracy $Q=0$, the density behaves as $Q^{3\beta_D/2}$ in the plane of centered energies and leads to a decay $t^{-(3\beta_D+2)}$ for even $\beta_D$. For the GOE, only the degenerate pair contributes. For IID levels, the GUE, and the GSE, this contribution is absent, and the triple degeneracy determines the decay. As a result,
\begin{align}
\big|K_S(t)-K_S^\infty\big|\,,\ \big|V_S(t)-V_S^\infty\big|\ \propto\ 
\begin{cases}
t^{-2}\,, & \beta_D=0,1\,,\\
t^{-8}\,, & \beta_D=2\,,\\
t^{-14}\,, & \beta_D=4\,.
\end{cases}
\label{eq:late}
\end{align}
For the GUE and the GSE, these power laws set in only once the deviation from the late-time value has become very small, and the curves in Fig.\,\ref{fig:late} are obtained from a high-precision evaluation of Eqs.\,\eqref{eq:KSavg} and \eqref{eq:VSavg}. We do not consider the entropy here, since $-p_n\ln p_n$ is not smooth where one of the probabilities vanishes.


\section{Conclusions and outlook}
\label{sec:disc}

We have shown how the spacing-ratio statistics enter the spreading of a state over the Krylov basis for three levels, where the spreading can be obtained in closed form. At infinite temperature, the Krylov basis depends on the spectrum only through the spacing ratio, while the scale of the spectrum sets the time scale. Since scale and shape are independent in the Gaussian ensembles, the averaged
spread complexity and Krylov variance reduce to one-dimensional integrals over the
spacing-ratio distribution. To our knowledge, these are the first exact results for the full time dependence of ensemble-averaged Krylov-space quantities in a setting in which correlations between different level spacings are present. They show, at a qualitative level, the growth, peak, and saturation of the spread complexity that have so far been observed numerically for large random
matrices or many-body Hamiltonians, where correlations between distant levels
also contribute.

The late-time values of the complexity and of the variance, as well as the temporal fluctuations of the complexity, do not depend on the spacing ratio and are the same for every three-level spectrum. The late-time entropy depends on the shape of the spectrum only weakly, through a function that is invariant under a duality exchanging nearly degenerate and nearly equally spaced configurations. The level statistics appear in the height of the peak, which increases with the Dyson index. They also appear in the exponents of the algebraic approach to the late-time values, and in the low-temperature limits of the ratios of the three quantities, which we obtained in closed form. The exponents coincide for IID levels and the GOE, for which the decay is $t^{-2}$. For $d$ levels, the same scale--shape decomposition holds in the Gaussian ensembles, and the scale follows a Gamma distribution with the parameter $\gamma_{\beta_D}=\beta_D d(d-1)/4+(d-1)/2$. The ensemble average of the spread complexity is then a $(d-2)$-dimensional integral over the shape of a finite sum of confluent hypergeometric functions, weighted by the joint distribution of $(d-2)$ consecutive spacing ratios \cite{Atas:2013gns}. For a single spectrum at infinite temperature, Eq.\,\eqref{eq:Xdef} gives $K_S(t)=\frac{d-1}{2}+\frac{2}{d}\sum_{k<l}X_{kl}\cos(\omega_{kl}t)$, where the coefficients $X_{kl}$ depend only on the shape. Writing $\omega_{kl}=\nu_{kl} \Lambda$ with the scale $\Lambda^2=\sum_k(E_k-\bar E)^2$, and averaging over $\Lambda$ as in Eq.\,\eqref{eq:cosavg}, one finds for the ensemble average
\begin{align}
K_S(t)=\frac{d-1}{2}+\frac{2}{d}\int\mathrm{d}^{d-2}r\,P_{\beta_D}(\mathbf{r})\sum_{k<l}X_{kl}(\mathbf{r})\;{}_1F_1\Big(\gamma_{\beta_D};\frac{1}{2};-\frac{\nu_{kl}^2(\mathbf{r})\,t^2}{d\beta_D}\Big)\,,
\label{eq:generald}
\end{align}
where $\mathbf{r}=(r_1,\dots,r_{d-2})$ are the consecutive spacing ratios with joint distribution $P_{\beta_D}(\mathbf{r})$ \cite{Atas:2013gns}, which reduces to Eq.\,\eqref{eq:atas} for $d=3$. For IID levels, $d\beta_D$ in the argument is replaced by $2$. For $d=2$, where no spacing ratio is present, one has $\gamma_{\beta_D} =\frac{1+\beta_D}{2}$, $X_{12}=-\frac{1}{2}$ and $\nu_{12}^2=2$, so that Eq.\,\eqref{eq:generald} gives
\begin{align}
K_S(t)=\frac{1}{2}\Big[1-{}_1F_1\Big(\frac{1+\beta_D}{2};\frac{1}{2};-\frac{t^2}{\beta_D}\Big)\Big]\,,
\label{eq:twolevel}
\end{align}
which agrees with the two-level result of Ref.\,\cite{Nandy:2024mml} up to the normalization of time, and $K_S(t)=\frac{1}{2}\big(1-e^{-t^2}\big)$ for IID levels \cite{Caputa:2024vrn}. For $d=3$, Eq.\,\eqref{eq:generald} reduces to Eq.\,\eqref{eq:main}, with $\frac{2}{3}X_{kl}=-\frac{\alpha_f}{9Q_r}$. The early growth $K_S(t)\simeq\langle b_1^2\rangle\,t^2$ follows for any $d$ from the Gamma distribution of the scale, with $\langle b_1^2\rangle=\frac{(d-1)(2+d\beta_D)}{d^2\beta_D}$ in the Gaussian ensembles and $\langle b_1^2\rangle=\frac{d-1}{d}$ for IID levels.

For $d\geq4$ levels at infinite temperature, the late-time complexity $K_S^{\infty}=\frac{d-1}{2}$ is again independent of the spectrum, but the Krylov basis depends on more than one shape variable, and $\delta K_S^2$ and $V_S^{\infty}$ are no longer universal. Evaluating Eq.\,\eqref{eq:generald} for $d=4$ would show how the peak of the averaged complexity evolves from two and three levels towards the large-$d$ limit. In this case, correlations between more distant levels, which determine the spectral rigidity and the ramp of the SFF, also contribute. It remains to be determined whether the peak of the averaged complexity is controlled by the correlations between neighboring spacings or by these long-range correlations. Ensembles in which short-range correlations are absent while weak long-range correlations persist \cite{Roy:2025mxt, Das:2025pwy} (see also Ref.\,\cite{Das:2023ass}) provide a setting in which the two can be distinguished. For three levels, the contribution of the correlation between adjacent spacings can be isolated by drawing the two spacings independently from their marginal distributions, which retains level repulsion but removes this correlation.

For large $d$, the exact integral representation may become impractical, since the number of frequencies grows as $d(d-1)/2$ and the coefficients depend on $(d-2)$ spacing ratios. It would be interesting to connect the exact results for small $d$ obtained here with the large-$d$ description in terms of the density of states and the SFF, for instance through the ensemble averages of the Lanczos coefficients for general $d$ \cite{Balasubramanian:2023kwd, Nandy:2024zcd}. For systems with a conserved charge, levels from different charge sectors do not repel each other, and the present approach could be used to study how this affects the total and the symmetry-resolved spread complexity \cite{Caputa:2025ozd}. We will return to some of these questions in future studies.


\section*{Acknowledgments} We would like to thank Pawel Caputa, Javier Magan and Alfonso V. Ramallo for useful discussions and comments on the draft. We thank the Nordic Institute for Theoretical Physics (Nordita) and the Instituto Galego de F\'isica de Altas Enerx\'ias (IGFAE) at the University of Santiago de Compostela for their hospitality during the final stages of this work. This work is supported by FWO-Vlaanderen projects G012222N and G0A2226N, and by the VUB Research Council through the Strategic Research Program High-Energy Physics. Claude Fable 5.1 and Opus 5.5 were used to optimize codes originally written by the author (in Mathematica 13.3), to generate Fig.\,\ref{fig:schematic}, and for grammatical corrections in the text. The author takes full responsibility for the scientific content and conclusions of this work.

\bibliographystyle{JHEP}
\bibliography{refs}

\end{document}